 \documentclass[final,3p]{elsarticle}
\usepackage{amssymb}
\journal{arXiv}
\usepackage{amsmath}
\usepackage{amssymb}
\usepackage{psfrag}
\usepackage{color}
\usepackage{stfloats}
\usepackage{fixltx2e}
\usepackage{mathrsfs}
\usepackage{tabularx}
\usepackage{booktabs}
\usepackage{colortbl}
\usepackage{rotating}
\usepackage{boldline}
\usepackage{bm}
\usepackage{changes}
\usepackage{url}
\usepackage{adjustbox}
\usepackage{graphicx}
\usepackage{cellspace}
\usepackage{lineno}

\begin{document}
%%%%%%%%%%%%%%%%%%%%%%%%%%%%%%%%%%%%%%%%%%%%%%%%%%%%%%%%%%%%%%%%%%%%%%%%
\renewcommand{\topfraction}{0.98}   % max fraction of floats at top
\renewcommand{\bottomfraction}{0.98}% max fraction of floats at bottom
\setcounter{topnumber}{3}
\setcounter{bottomnumber}{3}
\setcounter{totalnumber}{4}         % 2 may work better
\setcounter{dbltopnumber}{4}        % for 2-column pages
\renewcommand{\dbltopfraction}{0.98}% fit big float above 2-col. text
\renewcommand{\textfraction}{0.05}  % allow minimal text w. figs
\renewcommand{\floatpagefraction}{0.95}      % require fuller float pages
\renewcommand{\dblfloatpagefraction}{0.95}   % require fuller float pages
%%%%%%%%%%%%%%%%%%%%%%%%%%%%%%%%%%%%%%%%%%%%%%%%%%%%%%%%%%%%%%%%%%%%%%%%
%\newenvironment{workinprogress}{\par\color{red}}{\par}
\newcommand{\beq}{\begin{equation}}
\newcommand{\eeq}{\end{equation}}
\newcommand{\divg}{\mbox{\rm{div}}\,}
\newcommand{\Divg}{\mbox{\rm{Div}}\,}
\newcommand{\D}  {\displaystyle}
\newcommand{\DS} {\displaystyle}
\newcommand{\RM}[1]{\textit{\MakeUppercase{\romannumeral #1{}}}}
\newtheorem{remark}{\bf{{Remark}}}
\def\sca   #1{\mbox{\rm{#1}}{}}
\def\mat   #1{\mbox{\bf #1}{}}
\def\vec   #1{\mbox{\boldmath $#1$}{}}
\def\scas  #1{\mbox{{%\scriptsize
\footnotesize
{${\rm{#1}}$}}}{}}
\def\scaf  #1{\mbox{{\tiny{${\rm{#1}}$}}}{}}
\def\vecs  #1{\mbox{\boldmath{%\scriptsize
\footnotesize
{$#1$}}}{}}
\def\tens  #1{\mbox{\boldmath{%\scriptsize
\footnotesize
{$#1$}}}{}}
\def\tenf  #1{\mbox{{\sffamily{\bfseries {#1}}}}}
\def\ten   #1{\mbox{\boldmath $#1$}{}}
\def\Ass  {\overset{\hspace*{0.4cm} n_{\scas{el}}}
          {\underset{\scaf{c},\scaf{d}=1}{\msf{A}}}}
\def\ltr   #1{\mbox{\sffamily{#1}}}
\def\bltr  #1{\mbox{\sffamily{\bfseries{{#1}}}}}
\sloppy
%%%%%%%%%%%%%%%%%%%%%%%%%%%%%%%%%%%%%%%%%%%%%%%%%%%%%%%%%%%%%%%%%%%%%%%%
\begin{frontmatter}
%%%%%%%%%%%%%%%%%%%%%%%%%%%%%%%%%%%%%%%%%%%%%%%%%%%%%%%%%%%%%%%%%%%%%%%%
\title{\Large 
{\textsf{\textbf{Generative AI for material design:\\
A mechanics perspective from burgers to matter}}}}
%%%%%%%%%%%%%%%%%%%%%%%%%%%%%%%%%%%%%%%%%%%%%%%%%%%%%%%%%%%%%%%%%%%%%%%%
\author{{Vahidullah Tac}}
\author{{Ellen Kuhl}}%[orcid=0000-0002-6283-935X]
\address{Department of Mechanical Engineering, Stanford University, 
Stanford, California, United States\\[10.pt]
}
%%%%%%%%%%%%%%%%%%%%%%%%%%%%%%%%%%%%%%%%%%%%%%%%%%%%%%%%%%%%%%%%%%%%%%%%
% an effective abstract is brief and normally less than 200 words.
% abstracts must not exceed 250 words 
%%%%%%%%%%%%%%%%%%%%%%%%%%%%%%%%%%%%%%%%%%%%%%%%%%%%%%%%%%%%%%%%%%%%%%%%
\begin{abstract} % 
%%%%%%%%%%%%%%%%%%%%%%%%%%%%%%%%%%%%%%%%%%%%%%%%%%%%%%%%%%%%%%%%%%%%%%%%
Generative artificial intelligence 
offers a new paradigm to design matter in high-dimensional spaces.
However, its underlying mechanisms 
remain difficult to interpret 
and limit adoption in computational mechanics.
This gap is striking 
because its core tools--diffusion, stochastic differential equations, and inverse problems--are 
fundamental to the mechanics of materials.
Here we show that diffusion-based generative AI 
and computational mechanics are rooted in the same principles.
We illustrate this connection 
using a three-ingredient burger 
as a minimal benchmark for material design 
in a low-dimensional space, 
where both forward and reverse diffusion 
admit analytical solutions:
Markov chains with Bayesian inversion in the discrete case and 
the Ornstein–Uhlenbeck process with score-based reversal in the continuous case.
In both cases, 
forward diffusion adds noise to degrade structure, 
while reverse diffusion recovers structure from noise.
We extend this framework 
to a high-dimensional design space with 146 ingredients
and 8.9$\times 10^{43}$ possible configurations,
where analytical solutions become intractable. 
We therefore learn
the discrete and continuous reverse processes
using neural network models
that infer inverse dynamics from data.
We train the models on only 2,260 recipes and generate one million samples that capture the statistical structure of the data, including ingredient prevalence and quantitative composition.
We further generate five new burgers and validate them in a blinded restaurant-based sensory study with n = 101 participants, where three of the AI-designed burgers outperform the classical Big\,Mac in overall liking, flavor, and texture.
These results establish diffusion-based generative modeling as a physically grounded approach to design in high-dimensional spaces.
They position generative AI 
as a natural extension of computational mechanics, 
with applications from burgers to matter, 
and establish a path toward data-driven, 
physics-informed generative design.
%%%%%%%%%%%%%%%%%%%%%%%%%%%%%%%%%%%%%%%%%%%%%%%%%%%%%%%%%%%%%%%%%%%
Our source code, data, and examples are available at 
https:/\!/github.com/LivingMatterLab/AI4Food.
%%%%%%%%%%%%%%%%%%%%%%%%%%%%%%%%%%%%%%%%%%%%%%%%%%%%%%%%%%%%%%%%%%%
\end{abstract}
%%%%%%%%%%%%%%%%%%%%%%%%%%%%%%%%%%%%%%%%%%%%%%%%%%%%%%%%%%%%%%%%%%%%
%%%%%%%%%%%%%%%%%%%%%%%%%%%%%%%%%%%%%%%%%%%%%%%%%%%%%%%%%%%%%%%%%%%
\end{frontmatter}
%%%%%%%%%%%%%%%%%%%%%%%%%%%%%%%%%%%%%%%%%%%%%%%%%%%%%%%%%%%%%%%%%%%
%\end{document}
%\vspace*{0.5cm}
%%%%%%%%%%%%%%%%%%%%%%%%%%%%%%%%%%%%%%%%%%%%%%%%%%%%%%%%%%%%%%%%%%%%%%%%
\section{{\textsf{\textbf{Motivation}}}}
%%%%%%%%%%%%%%%%%%%%%%%%%%%%%%%%%%%%%%%%%%%%%%%%%%%%%%%%%%%%%%%%%%%%%%%%
\noindent
The modern hamburger emerged in the late 19th century 
as a simple combination of ground meat and bread \cite{smith2008}. 
Since then, 
its formulation has evolved through incremental variation; 
yet, its core design remains largely unchanged \cite{bohrer2019}.
Despite its global  
popularity--with more than 50 billion hamburgers 
consumed annually in the United States alone \cite{unruh16}--the
massive combinatorial space 
of possible ingredient configurations 
remains largely unexplored.
This challenge extends far beyond food: 
the design of matter and materials 
draws from vast libraries of candidate components 
that generate high-dimensional combinatorial spaces 
beyond systematic exploration \cite{jain2013}.
This gap between vast design potential 
and limited human search 
motivates a new approach to discovery.\\[6.pt] 
%%%%%%%%%%%%%%%%%%%%%%%%%%%%%%%%%%%%%%%%%%%%%%%%%%%%%%%%%%%%%%%%%%%%%%%%
{\textsf{\textbf{Generative AI for discovery.}}}
%%%%%%%%%%%%%%%%%%%%%%%%%%%%%%%%%%%%%%%%%%%%%%%%%%%%%%%%%%%%%%%%%%%%%%%%
Generative artificial intelligence is redefining how we explore high-dimensional design spaces and discover new materials across science and engineering \cite{goodfellow2014,kingma2014}. 
Scientists now apply these methods across domains such as 
computational drug design \cite{jumper2021}, 
molecular generation \cite{butler2018}, and 
materials discovery \cite{zeni2025}, where they propose new candidates that extend far beyond existing datasets \cite{gomez2018}. 
These problems share a fundamental challenge: the underlying design spaces grow combinatorially, while available training data remain sparse \cite{sohl2015}. 
Diffusion-based generative models address this challenge by introducing a forward process that progressively randomizes data and a reverse process that reconstructs structure from noise \cite{ho2020}. 
This forward--reverse formulation enables controlled sampling of complex distributions and produces new candidates that remain consistent with observed data, while exploring previously unseen regions of the design space \cite{song2021}. 
Diffusion models were originally derived from non-equilibrium statistical physics \cite{sohl2015}. 
Our objective is {\it{not}} to re-establish this origin, but to translate these concepts into the language of computational mechanics. This perspective casts generative modeling in terms of stochastic processes, transport, and inverse problems, and makes these methods directly accessible to the mechanics community. \\[6.pt]
%%%%%%%%%%%%%%%%%%%%%%%%%%%%%%%%%%%%%%%%%%%%%%%%%%%%%%%%%%%%%%%%%%%%%%%%
{\textsf{\textbf{A mechanics perspective.}}}
%%%%%%%%%%%%%%%%%%%%%%%%%%%%%%%%%%%%%%%%%%%%%%%%%%%%%%%%%%%%%%%%%%%%%%%%
From a mechanics perspective, diffusion models take a familiar form: stochastic dynamical systems governed by drift--diffusion equations in high-dimensional state spaces \cite{risken1996}. 
The forward process adopts {\it{Fokker--Planck diffusion}} 
of probability densities \cite{fokker1914,planck1917},
${\partial p}/{\partial t}
= -\nabla \cdot \vec{j}$,
where the flux,
$\vec{j} 
= p \, \vec{b}
- \frac{1}{2} \, \ten{D} \cdot \nabla p$,
governs how 
the probability density $p$ 
drifts and diffuses under 
the velocity $\vec{b}$ that controls deterministic transport, and
the diffusion tensor $\ten{D}$ that controls stochastic spreading.
Without the drift, % and for deterministic spreading, 
this reduces to classical diffusion 
along concentration gradients \cite{gardiner2009},
similar to
{\it{Fickian diffusion}} of concentrations \cite{fick1855} or 
{\it{Fourier diffusion}} of heat \cite{fourier1822},
${\partial p}/{\partial t}
= -\nabla \cdot \vec{j}$,
where the mass or heat flux,
$\vec{j} 
= - \ten{D} \cdot \nabla p$,
governs how 
concentration or heat $p$
diffuse under  
the diffusivity or conductivity tensor $\ten{D}$.
In the discrete setting, 
the same evolution reduces to a random walk on a graph, 
${\mathrm d p_i}/{\mathrm d t} = \sum_{j} {\textsf{L}}_{ij} \, p_j$,
where ${\textsf{L}}$ denotes the {\it{graph Laplacian}}   
or {\it{Kirchhoff matrix}} \cite{kirchhoff1847}, 
which governs the diffusion of probability mass 
across neighboring nodes 
$p_i$ and $p_j$ of the graph \cite{chung1997}. 
The reverse process transports probability mass 
up density gradients through a learned drift term 
and parallels driven transport in phase-separating systems 
such as {\it{Cahn--Hilliard diffusion}} \cite{cahn1958}. 
The flux of these systems, $\vec{j} = -\ten{D} \cdot \nabla \mu$,
derives from {\it{Onsager's variational principle}} \cite{onsager1931},
which leads to linear flux-force relations 
driven by gradients of the chemical potential,
$\mu = \delta \mathcal{F} / \delta p$, 
rather than by concentration gradients $\nabla p$ alone, 
to enable unmixing and uphill transport.
This perspective places generative diffusion models within the well-established framework of gradient flows, variational principles, and entropy-driven evolution, and highlights that the mathematical tools of computational mechanics already provide a natural language to analyze, reinterpret, and extend diffusion models in generative AI \cite{ambrosio2008,jordan1998,tac2026,villani2009}.\\[6.pt]
%%%%%%%%%%%%%%%%%%%%%%%%%%%%%%%%%%%%%%%%%%%%%%%%%%%%%%%%%%%%%%%%%%%%%%%%
{\textsf{\textbf{Food as a model system.}}}
%%%%%%%%%%%%%%%%%%%%%%%%%%%%%%%%%%%%%%%%%%%%%%%%%%%%%%%%%%%%%%%%%%%%%%%%
Food represents a high-impact societal challenge for generative AI, 
as it forms a class of complex, multicomponent materials 
in which composition, structure, and processing 
jointly determine mechanical response and sensory perception \cite{datta2025,figura2023,li2026,stpierre2024}. 
Food scientists are beginning to use machine learning to capture relationships between ingredients, texture, and consumer preference, which enables data-driven optimization of food products \cite{alsarayreh2023,datta2022,gunning2026,tagkopoulos2022,vandenbedem2026,zohdi2024}. 
Generative models extend this approach by proposing entirely new formulations, which opens the door to systematically explore of the vast combinatorial ingredient space \cite{kuhl2025}. 
A new generation of AI-driven food technology companies, including pioneers such as NotCo, is already translating these ideas into practice by designing plant-based foods that replicate the sensory properties of animal products \cite{notco2021a,notco2021b}. 
These developments position food as an ideal testbed for generative design in high-dimensional material systems \cite{tac2026}.\\[6.pt]
%%%%%%%%%%%%%%%%%%%%%%%%%%%%%%%%%%%%%%%%%%%%%%%%%%%%%%%%%%%%%%%%%%%%%%%%
{\textsf{\textbf{Burgers as a benchmark.}}}
%%%%%%%%%%%%%%%%%%%%%%%%%%%%%%%%%%%%%%%%%%%%%%%%%%%%%%%%%%%%%%%%%%%%%%%%
To illustrate these ideas, we introduce burgers as a minimal yet expressive model system for generative design. We begin with a three-ingredient problem, in which each burger consists of bun, patty, and cheese, and train diffusion models on only two examples, hamburger and cheeseburger, to illustrate how forward and reverse diffusion generate new combinations in a finite state space. 
From a mechanics perspective, this system defines a {\it{low-dimensional}} phase space in which we can explicitly analyze and illustrate discrete and continuous diffusion processes, and use it as a transparent setting to study transport, entropy, and structure formation.
We then extend this framework to a realistic setting with 146 ingredients and 2,260 training recipes, where the design space becomes astronomically large and cannot be explored exhaustively. 
Within this {\it{high-dimensional}} setting, diffusion models enable principled sampling of novel burgers that remain consistent with the training data, while exploring new regions of ingredient space. This progression--from a simple discrete system to a complex continuous problem--establishes burgers as a canonical model for generative design and provides a concrete entry point for applying the tools of computational mechanics to high-dimensional discovery problems.\\[6.pt]
%%%%%%%%%%%%%%%%%%%%%%%%%%%%%%%%%%%%%%%%%%%%%%%%%%%%%%%%%%%%%%%%%%%%%%%%
{\textsf{\textbf{Outline.}}}
%%%%%%%%%%%%%%%%%%%%%%%%%%%%%%%%%%%%%%%%%%%%%%%%%%%%%%%%%%%%%%%%%%%%%%%%
Our objective is to establish diffusion models for generative material design, first through a minimal benchmark problem, a three-ingredient burger for which analytical solutions are possible, and then through a real-world problem with more than a hundred possible ingredients for which generative modeling becomes imperative. 
Section~\ref{sec02} introduces {\it{discrete diffusion}} for ingredient selection using the three-ingredient burger model system, characterized by the discrete ingredient vector 
$\vec{x} = [\; x_\text{bun}, x_\text{patty}, x_\text{cheese} \;] \in \{0,1\}^3$.
Section~\ref{sec03} extends this formulation to {\it{continuous diffusion}} for ingredient quantification using the same three-ingredient model system, now characterized by the continuous weight vector 
$\vec{w} = [\; w_\text{bun}, w_\text{patty}, w_\text{cheese} \;] \in \mathbb{R}^3$.
Section~\ref{sec04} generalizes both formulations from the illustrative three-ingredient space to a real-world 146-ingredient space and discusses the challenges associated with high-dimensional generative material design.
All three sections share a common structure: they begin with {\it{forward diffusion}}, which transforms structure into noise, proceed with {\it{reverse diffusion}}, which reconstructs structure from noise, highlight {\it{illustrative examples}} of forward and reverse diffusion, and conclude with the {\it{generation of new designs}} by sampling from the diffusion model. 
Section~\ref{sec05} concludes with a discussion of the results and outlines key challenges and opportunities for computational mechanics in generative AI for materials.
%%%%%%%%%%%%%%%%%%%%%%%%%%%%%%%%%%%%%%%%%%%%%%%%%%%%%%%%%%%%%%%%%%%%%%%%%
%%%%%%%%%%%%%%%%%%%%%%%%%%%%%%%%%%%%%%%%%%%%%%%%%%%%%%%%%%%%%%%%%%%%%%%%%%
\section{{\textsf{\textbf{Discrete diffusion}}}}\label{sec02}
\begin{figure}[h]
\centering
\includegraphics[width=\textwidth]{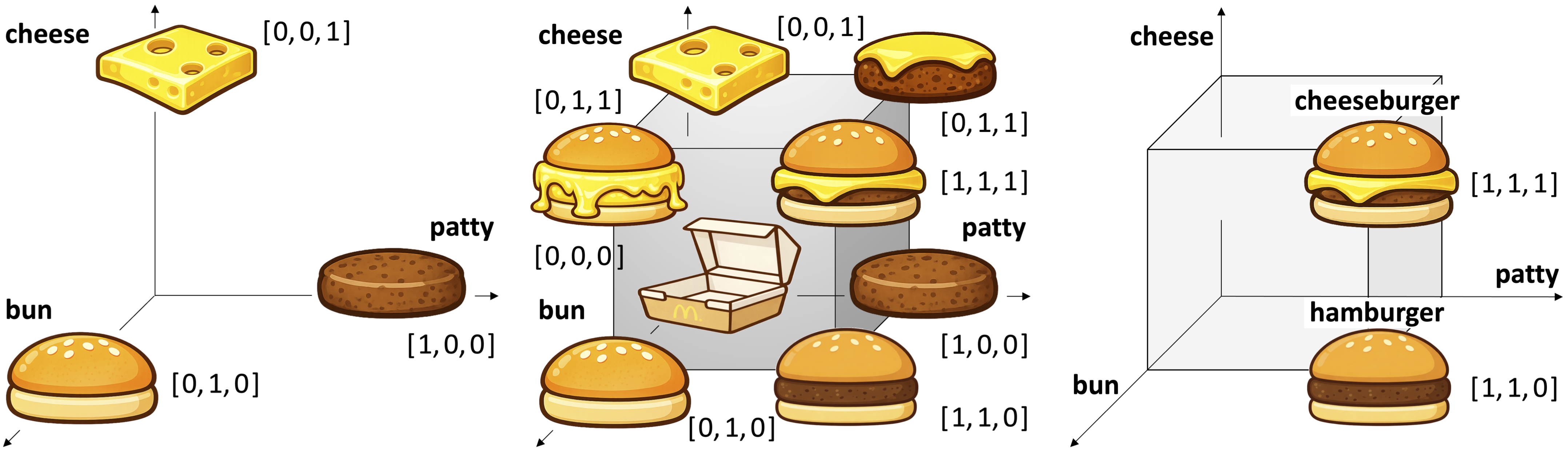} 
\caption{{\bf{\sffamily{Three-ingredient burger problem.}}} 
Three-ingredient space 
$[\, x_\text{bun}, x_\text{patty}, x_\text{cheese} \;]$
with patty, bun, and cheese (left);
with each ingredient either present or absent,
$\vec{x} = [\, x_\text{bun}, x_\text{patty}, x_\text{cheese}\, ] 
\in  \{0,1\}^3$,
generating $2^3=8$ eight possible burgers (middle);
training data with
cheeseburger $\vec{x}_1 = [\,1,1,1\,]$ and
hamburger    $\vec{x}_2 = [\,1,1,0\,]$ (right).}
\label{fig01}
\end{figure}

%%%%%%%%%%%%%%%%%%%%%%%%%%%%%%%%%%%%%%%%%%%%%%%%%%%%%%%%%%%%%%%%%%%%%%%%%%
\noindent
We illustrate the concept of {\it{discrete diffusion}} 
through a minimal benchmark problem: 
{\it{ingredient selection}} for a three-ingredient burger.
For this example, we introduce 
three binary ingredients,
bun, patty, and cheese,
that are either present or not, 
\beq
\vec{x} = [\, x_{\text{bun}}, x_{\text{patty}}, x_{\text{cheese}} \,]
\; \in \; \{0,1\}^3 \,,
\eeq
Combinatorics introduces 
a total of 2$^3$ = 8 possible burgers,
a bun with patty with cheese, 
a bun with a patty, a bun with cheese, 
a patty with cheese, 
just a bun, patty, or cheese,
or no ingredients at all (Fig. \ref{fig01}). \\[6.pt]
%%%%%%%%%%%%%%%%%%%%%%%%%%%%%%%%%%%%%%%%%%%%%%%%%%%%%%%%%%%%%%%%%%%%%%%%%%
{\textsf{\textbf{Training data.}}}
%%%%%%%%%%%%%%%%%%%%%%%%%%%%%%%%%%%%%%%%%%%%%%%%%%%%%%%%%%%%%%%%%%%%%%%%%%
We assume the training data %dataset 
contain two burgers,
\beq
\vec{x}_{1} = [\,1,1,1\,] \quad \mbox{and} \quad
\vec{x}_{2} = [\,1,1,0\,],
\eeq
a cheeseburger, i.e., a bun with patty with cheese, and
a hamburger, i.e., a bun with patty. 
We assume that both burgers are 
equally present with an empirical data distribution,
\beq
  p_{\text{data}}(\vec{x})
= \mbox{$\frac{1}{2}$} \, \delta(\vec{x} - \vec{x}_1)
+ \mbox{$\frac{1}{2}$} \, \delta(\vec{x} - \vec{x}_2) \,,
\eeq
where 
$\delta$ denotes the Kronecker delta, such that
$p_{\text{data}}(\vec{x}_1) = \frac{1}{2}$ and
$p_{\text{data}}(\vec{x}_2) = \frac{1}{2}$ and
$p_{\text{data}}(\vec{x}) = 0$ otherwise. \\[6.pt]
%%%%%%%%%%%%%%%%%%%%%%%%%%%%%%%%%%%%%%%%%%%%%%%%%%%%%%%%%%%%%%%%%%%%%%%%%%
{\textsf{\textbf{Structure.}}}
%%%%%%%%%%%%%%%%%%%%%%%%%%%%%%%%%%%%%%%%%%%%%%%%%%%%%%%%%%%%%%%%%%%%%%%%%%
From the empirical data distribution $p_{\text{data}}(\vec{x})$, 
we compute the marginal probabilities ${\textsf{P}}$ of each ingredient,
\beq
{\textsf{P}}(x_{\text{bun}}=1)  =1   \quad \mbox{and} \quad
{\textsf{P}}(x_{\text{patty}}=1)=1  \quad \mbox{and} \quad
{\textsf{P}}(x_{\text{cheese}}=1)= \mbox{$\frac{1}{2}$}.
\eeq
This means that, under the data distribution,
the marginal probability that the bun is present is one,
that the patty is present is one, and
that the cheese is present is one half:
Bun and patty are {\it{deterministic}},
while cheese is {\it{stochastic}}.\\[6.pt]
%%%%%%%%%%%%%%%%%%%%%%%%%%%%%%%%%%%%%%%%%%%%%%%%%%%%%%%%%%%%%%%%%%%%%%%%%%
{\textsf{\textbf{Entropy.}}}
%%%%%%%%%%%%%%%%%%%%%%%%%%%%%%%%%%%%%%%%%%%%%%%%%%%%%%%%%%%%%%%%%%%%%%%%%%
The {\it{Shannon entropy}} 
quantifies the uncertainty associated 
with the probability distribution $p$ \cite{shannon1948},
\beq
H(p) = - \mbox{$\sum_{\vecs{x}}$} \, p(\vec{x})\log p(\vec{x}).
\eeq
For the empirical data distribution, 
the probability mass is equally split
between two states,
the bun with patty with cheese,
$p_{\text{data}}(\vec{x}_1) = \frac{1}{2}$ and
the bun with patty,
$p_{\text{data}}(\vec{x}_2) = \frac{1}{2}$,
thus,
\beq
  H(p_{\text{data}})
= -2 \; [\, \, \mbox{$\frac{1}{2}$} \,
  \log (\mbox{$\frac{1}{2}$}) \;]
= \log (2).
\eeq
This entropy reflects the fact that the dataset contains
uncertainty along only one coordinate, cheese,
while two other coordinates, bun and patty, are fixed.
If all eight states were equally likely, 
$p_{\text{data}} (\vec{x}) =\frac{1}{8}$ for all $\vec{x}$,
maximum entropy would occur, with
\beq
  H_{\text{max}}
= -8 \; [\, \, \mbox{$\frac{1}{8}$} \,
  \log (\mbox{$\frac{1}{8}$}) \;]
= \log (8).
\eeq
This implies that 
the data distribution has a significantly lower entropy
than the uniform distribution,
$H(p_{\text{data}}) < H_{\text{max}}$.
Forward diffusion will gradually increase entropy,
and drive the distribution from $\log (2)$ toward $\log (8)$. %\\[6.pt]
%%%%%%%%%%%%%%%%%%%%%%%%%%%%%%%%%%%%%%%%%%%%%%%%%%%%%%%%%%%%%%%%%%%%%%%%
 \\[6.pt]
%%%%%%%%%%%%%%%%%%%%%%%%%%%%%%%%%%%%%%%%%%%%%%%%%%%%%%%%%%%%%%%%%%%%%%%%%%
{\it{{\textsf{\textbf{Geometric interpretation.}}}
%%%%%%%%%%%%%%%%%%%%%%%%%%%%%%%%%%%%%%%%%%%%%%%%%%%%%%%%%%%%%%%%%%%%%%%%%%
The three binary ingredients 
define a three-dimensional discrete state space $\{0,1\}^3$, 
and the eight possible burgers 
correspond to the vertices of the unit cube $[\,0,1\,]^3$. 
In this discrete space,
probability mass is initially concentrated 
on two adjacent vertices of a single edge,
rather than spread across all eight vertices of the cube.
Although the state space is three-dimensional,
the dataset varies only along one coordinate, cheese,
while the other two coordinates, bun and patty, are fixed. 
Forward diffusion will gradually move the probability mass
from these two vertices 
towards all other vertices of the cube (Fig. \ref{fig01}).}}
%%%%%%%%%%%%%%%%%%%%%%%%%%%%%%%%%%%%%%%%%%%%%%%%%%%%%%%%%%%%%%%%%%%%%%%%%%
\subsection{\textsf{\textbf{Forward diffusion}}}
\begin{figure}[h]
\centering
\includegraphics[width=\textwidth]{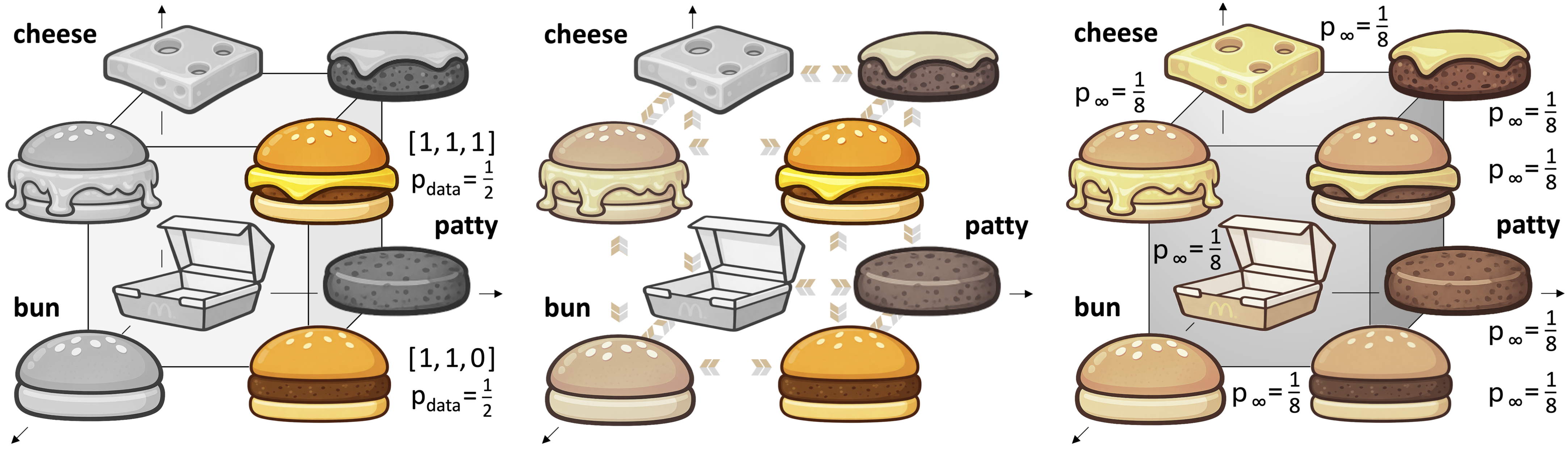} 
\caption{{\bf{\sffamily{Discrete modeling of forward diffusion.}}} 
Three-ingredient space with eight possible burgers 
with training data, 
cheeseburger $\vec{x}_1 = [\,1,1,1\,]$ and
hamburger    $\vec{x}_2 = [\,1,1,0\,]$  
with equal probabilities
$p_{\text{data}}(\vec{x}_1) = 0.50$ and
$p_{\text{data}}(\vec{x}_2) = 0.50$, highlighted in color
(left);
forward diffusion with independent equal-probabilty flipping of the three ingredients gradually diffuses probabilities across the cube (middle);
converged state of maximum entropy with equal probabilities 
$p_{\infty} =\frac{1}{8}$
across all eight burgers (right).}
\label{fig02}
\end{figure}

%%%%%%%%%%%%%%%%%%%%%%%%%%%%%%%%%%%%%%%%%%%%%%%%%%%%%%%%%%%%%%%%%%%%%%%%%%
\noindent
The forward diffusion process gradually destroys structure
by randomly perturbing each ingredient.
Specifically, at each time step, every ingredient,
$\{ \, \text{bun, patty, cheese} \, \}$, 
flips independently with a probability $\beta$.
A flip means that a present ingredient becomes absent,
or an absent ingredient becomes present (Fig. \ref{fig02}).\\[6.pt]
%%%%%%%%%%%%%%%%%%%%%%%%%%%%%%%%%%%%%%%%%%%%%%%%%%%%%%%%%%%%%%%%%%%%%%%%%%
{\textsf{\textbf{Single-ingredient transition.}}}
%%%%%%%%%%%%%%%%%%%%%%%%%%%%%%%%%%%%%%%%%%%%%%%%%%%%%%%%%%%%%%%%%%%%%%%%%%
For a single binary ingredient $x_i \in \{0,1\}$,
the transition matrix $\mat{Q}_1$ 
governs the transition
%$\vec{x}_{t} = \mat{Q}_1 \cdot \vec{x}_{t-1}$,
from one time step to the next, 
\beq
\mat{Q}_1 =
\left[ \; 
\begin{array}{c@{\hspace*{0.1cm}}c}
[\,1-\beta\,] & \beta \\
\beta   & [\, 1-\beta \,]
\end{array}
\; \right] \,.
\eeq
Here, $\beta \in (0,1)$ is the per-ingredient flip probability,
which controls the strength of the diffusion level.
With a probability $[\,1-\beta\,]$, the ingredient remains unchanged;
with a probability $\beta$, the ingredient flips.
For small $\beta$, ingredients flip slowly, structure decays gradually, and mixing is slow;
for large $\beta$, ingredients flip frequently, structure disappears quickly, and the ingredient list converges rapidly to a uniform distribution. \\[6.pt]
%%%%%%%%%%%%%%%%%%%%%%%%%%%%%%%%%%%%%%%%%%%%%%%%%%%%%%%%%%%%%%%%%%%%%%%%%%
{\textsf{\textbf{All-ingredient transition.}}}
%%%%%%%%%%%%%%%%%%%%%%%%%%%%%%%%%%%%%%%%%%%%%%%%%%%%%%%%%%%%%%%%%%%%%%%%%%
Because all three ingredients flip {\it{independently}},
the full transition matrix for the burger is the tensor product,
\beq
\mat{Q}_3 = \mat{Q}_1 \otimes \mat{Q}_1 \otimes \mat{Q}_1 \,.
\eeq
The matrix $\mat{Q}_3$ 
defines a {\it{Markov chain}} 
on the eight vertices of the cube $\{0,1\}^3$.
Here we model diffusion through a discrete-time Markov chain, 
because it defines a simple stochastic process 
on a finite state space 
with analytically tractable forward transitions 
and exact reverse dynamics via Bayes' theorem.
We define the forward diffusion process 
through a transition kernel
which independently flips each ingredient with probability $\beta$,
\beq
  p(\vec{x}_t \,|\, \vec{x}_{t-1})
= \mbox{$\prod_{i=1}^{3}$} 
[\,\beta\, [\,1-\delta_{\vecs{x}_{t,i},\vecs{x}_{t-1,i}}\,]
   [\, 1-\beta \,] \,\delta_{\vecs{x}_{t,i},\vecs{x}_{t-1,i}} \,] \,,
\label{forward_kernel}
\eeq
where $\delta$ 
denotes the Kronecker delta, 
which equals one if the ingredient state remains unchanged 
and zero if it flips. \\[6.pt]
%%%%%%%%%%%%%%%%%%%%%%%%%%%%%%%%%%%%%%%%%%%%%%%%%%%%%%%%%%%%%%%%%%%%%%%%%%
{\textsf{\textbf{One-step transition probability.}}}
%%%%%%%%%%%%%%%%%%%%%%%%%%%%%%%%%%%%%%%%%%%%%%%%%%%%%%%%%%%%%%%%%%%%%%%%%%
We can introduce 
the explicit one-step transition probability 
of the forward diffusion Markov chain (\ref{forward_kernel}),
\beq
   p(\vec{y} \mid \vec{x})
=  \beta^{d(\vecs{x},\vecs{y})} 
\, [\, 1-\beta \,]^{3-d(\vecs{x},\vecs{y})},
\label{discrete_one_step}
\eeq
where $d(\vec{x},\vec{y})$ is the {\it{Hamming distance}},
the number of ingredients by which two burgers $\vec{x}$ and $\vec{y}$ differ. 
For a three-ingredient burger, within one diffusion step,
the probabilities of no flip, one flip, two flips, and all three flips are
\beq
{\textsf{P}}(d=0 |\, \vec{x}) = [\,1-\beta\,]^3 \quad
{\textsf{P}}(d=1 |\, \vec{x}) = 3\beta \,[\,1-\beta\,]^2 \quad
{\textsf{P}}(d=2 |\, \vec{x}) = 3\beta^2 [\,1-\beta\,] \quad
{\textsf{P}}(d=3 |\, \vec{x}) = \beta^3 \,.
\eeq
For the example of a moderate flip probability of $\beta$=0.025, the flip probabilities would be
${\textsf{P}}(d=0 |\, \vec{x})$ = 0.92686 and
${\textsf{P}}(d=1 |\, \vec{x})$ = 0.07130 and
${\textsf{P}}(d=2 |\, \vec{x})$ = 0.00183 and
${\textsf{P}}(d=3 |\, \vec{x})$ = 0.00002.
This implies that, 
in one step, 
at a probability of 0.92686,
the burger is usually preserved, 
but occasionally loses or gains ingredients.
Over repeated steps, these random flips accumulate.\\[6.pt]
%%%%%%%%%%%%%%%%%%%%%%%%%%%%%%%%%%%%%%%%%%%%%%%%%%%%%%%%%%%%%%%%%%%%%%%%%%
{\textsf{\textbf{Evolution of probability distribution.}}}
%%%%%%%%%%%%%%%%%%%%%%%%%%%%%%%%%%%%%%%%%%%%%%%%%%%%%%%%%%%%%%%%%%%%%%%%%%
To calculate the evolution of the distribution, 
we introduce $p_t(\vec{x})$, the probability distribution at time $t$,
which evolves according to
\beq
   p_{t+1}(\vec{y})
= \mbox{$\sum_{\vecs{x}}$} \, p_t(\vec{x})\,p(\vec{y} \,|\, \vec{x}).
\label{evolution_p}
\eeq
This equation expresses the probability of burger $\vec{y}$ at time $(t+1)$ 
as a sum over all possible previous burgers $\vec{x}$, 
weighted by their probability $p_t(\vec{x})$
and the transition probability $p(\vec{y} \mid \vec{x})$.
It defines a discrete conservation law for probability mass on the state space,
analogous to transport equations in continuum mechanics, 
and describes the evolution of the distribution at the ensemble level.\\[6.pt]
%%%%%%%%%%%%%%%%%%%%%%%%%%%%%%%%%%%%%%%%%%%%%%%%%%%%%%%%%%%%%%%%%%%%%%%%%%
{\textsf{\textbf{Analytical solution.}}} 
%%%%%%%%%%%%%%%%%%%%%%%%%%%%%%%%%%%%%%%%%%%%%%%%%%%%%%%%%%%%%%%%%%%%%%%%%%
The forward diffusion process admits a closed-form solution. 
For a general initial distribution $p_0(\vec{x})$, 
the solution follows by linear superposition,
$ p_t(\vec{x}) 
= \sum_{\vecs{x}_0} 
  p_0(\vec{x}_0)\, 
  p_t(\vec{x} \mid \vec{x}_0)$.
%and converges to the uniform distribution as $t \to \infty$.
For an initial configuration $\vec{x}_0$, 
like in our three-ingredient burger problem, 
the distribution at time $t$ depends only on the Hamming distance 
$d(\vec{x}_0,\vec{x})$ and takes the following explicit form,
\beq
  p_t(\vec{x} \mid \vec{x}_0) 
= q_t^{\,d(\vecs{x}_0,\vecs{x})} 
  [\,1 - q_t\,]^{3-d(\vecs{x}_0,\vecs{x})}
  \qquad \mbox{with} \qquad
  q_t = \tfrac{1}{2} \, [\, 1 - [\, 1 - 2\beta \,]^t] \,,
  \label{evolution_hamming}
\eeq
where $q_t$ denotes the cumulative flip probability after $t$ steps. 
This implies that in practice, 
we can simulate the evolution of probabilities in equation (\ref{evolution_p})
simply by sampling trajectories of the underlying Markov chain, 
which generates stochastic paths 
from $\vec{x}_0 \to \vec{x}_1 \to ... \to \vec{x}_T$ 
through independent Bernoulli updates at each time step. 
%\\[6.pt]
%%%%%%%%%%%%%%%%%%%%%%%%%%%%%%%%%%%%%%%%%%%%%%%%%%%%%%%%%%%%%%%%%%%%%%%%
 \\[6.pt]
%%%%%%%%%%%%%%%%%%%%%%%%%%%%%%%%%%%%%%%%%%%%%%%%%%%%%%%%%%%%%%%%%%%%%%%%
%%%%%%%%%%%%%%%%%%%%%%%%%%%%%%%%%%%%%%%%%%%%%%%%%%%%%%%%%%%%%%%%%%%%%%%%%%
{\it{{\textsf{\textbf{Geometric interpretation.}}}
%%%%%%%%%%%%%%%%%%%%%%%%%%%%%%%%%%%%%%%%%%%%%%%%%%%%%%%%%%%%%%%%%%%%%%%%%%
Forward diffusion is a symmetric random walk on the cube graph.
The eight burgers represent the vertices of the cube.
The edges connect burgers that differ in exactly one ingredient.
Probability flows along the edges of the cube. %with equal weight.
Because transitions are symmetric, no burger is preferred.
This process gradually spreads probability mass
from the two training vertices to all eight vertices of the cube.
Over time, this mixing destroys the original structure
that the bun and patty are always present
and the distribution approaches uniformity (Fig. \ref{fig02}).}}\\[6.pt]
%%%%%%%%%%%%%%%%%%%%%%%%%%%%%%%%%%%%%%%%%%%%%%%%%%%%%%%%%%%%%%%%%%%%%%%%%%
{\textsf{\textbf{Entropy and convergence.}}}
%%%%%%%%%%%%%%%%%%%%%%%%%%%%%%%%%%%%%%%%%%%%%%%%%%%%%%%%%%%%%%%%%%%%%%%%%%
For $0<\beta<1$, 
every ingredient has 
a positive probability of flipping $\beta$, and
a positive probability of remaining unchanged $[\,1-\beta\,]$.
Consequently, 
from any burger state, 
it is possible to reach any other burger state 
in a finite number of steps.
The Markov chain defined by $\mat{Q}_3$ is therefore {\it{irreducible}}.
Because self-transitions occur with probability $[\, 1-\beta\,]^3>0$,
the chain is also {\it{aperiodic}}.
Hence, the chain is {\it{ergodic}} 
and admits a unique stationary distribution. \\[6.pt]
%%%%%%%%%%%%%%%%%%%%%%%%%%%%%%%%%%%%%%%%%%%%%%%%%%%%%%%%%%%%%%%%%%%%%%%%%%
{\textsf{\textbf{Stationary distribution.}}}
%%%%%%%%%%%%%%%%%%%%%%%%%%%%%%%%%%%%%%%%%%%%%%%%%%%%%%%%%%%%%%%%%%%%%%%%%%
Since the transition matrix $\mat{Q}_3$ is symmetric,
all states are treated equally.
The stationary distribution must therefore be {\it{uniform}},
\beq
  p_\infty(\vec{x})
= \mbox{$\frac{1}{8}$}
  \qquad
  \vec{x} \in \{0,1\}^3.
\eeq
Regardless of the initial distribution,
as $t \to \infty$,
the stationary distribution will converge
to this uniform distribution,
$p_t \to p_\infty$.
In other words, 
repeated random ingredient flips 
eventually make all eight burgers equally likely.\\[6.pt]
%%%%%%%%%%%%%%%%%%%%%%%%%%%%%%%%%%%%%%%%%%%%%%%%%%%%%%%%%%%%%%%%%%%%%%%%%%
{\textsf{\textbf{Entropy evolution.}}}
%%%%%%%%%%%%%%%%%%%%%%%%%%%%%%%%%%%%%%%%%%%%%%%%%%%%%%%%%%%%%%%%%%%%%%%%%%
Recall that the Shannon entropy is
\beq
H(p) = - \mbox{$\sum_{\vecs{x}}$} \, p(\vec{x}) \, \log (p(\vec{x})) \,.
\eeq
Initially, the data distribution is supported on two burgers.
At stationarity, all eight burgers are equally likely.
For the entropy, this implies,
\beq
H(p_0) =\log (2)
\qquad \mbox{and} \qquad
H(p_\infty) =\log (8) = 3 \, \log (2) =  H_{\text{max}}.
\eeq
The entropy increases from $\log (2)$ to $\log (8)$,
a threefold increase in uncertainty.
Forward diffusion increases entropy
and converges to the maximum-entropy distribution.  \\[6.pt]
%%%%%%%%%%%%%%%%%%%%%%%%%%%%%%%%%%%%%%%%%%%%%%%%%%%%%%%%%%%%%%%%%%%%%%%%%%
{\it{{\textsf{\textbf{Burger interpretation.}}}
%%%%%%%%%%%%%%%%%%%%%%%%%%%%%%%%%%%%%%%%%%%%%%%%%%%%%%%%%%%%%%%%%%%%%%%%%%
Initially, 
bun and patty were deterministic, 
and only cheese varied, $x_{\mathrm{cheese}}=\{0,1\}$,
After sufficient diffusion,
bun may be present or absent, $x_{\mathrm{bun}}=\{0,1\}$,
patty may be present or absent, $x_{\mathrm{patty}}=\{0,1\}$,
cheese may be present or absent, $x_{\mathrm{cheese}}=\{0,1\}$.
Forward diffusion forgets structure.
All combinations become equally likely.
The structured dataset dissolves into complete uncertainty.
Forward diffusion spreads the probability mass
from the two initial vertices
$\vec{x}_1 = [\, 1,1,1 \,]$ and $\vec{x}_2 = [\, 1,1,0 \,]$,
across the cube, and in the long-time limit,
converges to the uniform distribution on all eight vertices 
(Fig. \ref{fig02}).}}\\[6.pt]
%%%%%%%%%%%%%%%%%%%%%%%%%%%%%%%%%%%%%%%%%%%%%%%%%%%%%%%%%%%%%%%%%%%%%%%%%%
{\it{\textsf{\textbf{Analogy to mechanics.}}}
%%%%%%%%%%%%%%%%%%%%%%%%%%%%%%%%%%%%%%%%%%%%%%%%%%%%%%%%%%%%%%%%%%%%%%%%%%
We can interpret the discrete forward diffusion process 
as a random walk on a finite state space, 
governed by the Markov update,
$p_{t+1}(\vec{y}) = \sum_{\vecs{x}} p_t(\vec{x})\, P(\vec{y} \to \vec{x})$,
which is formally analogous 
to a master equation that describes stochastic hopping 
between discrete configurations in lattice-based models. 
In this interpretation, the transition probabilities $P(\vec{y} \to \vec{x})$ define local hopping rates between neighboring states. 
The transition kernel 
defines a stochastic nearest-neighbor interaction on the cube, 
where each ingredient behaves 
like a binary degree of freedom 
that switches between two states with flip probability $\beta$.
Repeated application of this operator 
leads to a progressive spreading of probability mass, 
and over time, 
the distribution converges toward a uniform equilibrium 
that corresponds to a system 
with no energetic preference among configurations.
In this formulation, the generator $\textsf{L}$ 
acts as a discrete Laplacian on the state graph, 
and the evolution corresponds to diffusion of probability mass 
across neighboring configurations \cite{schafer2020}.}
%%%%%%%%%%%%%%%%%%%%%%%%%%%%%%%%%%%%%%%%%%%%%%%%%%%%%%%%%%%%%%%%%%%%%%%%%%
\subsection{\textsf{\textbf{Reverse diffusion}}}
\begin{figure}[h]
\centering
\includegraphics[width=\textwidth]{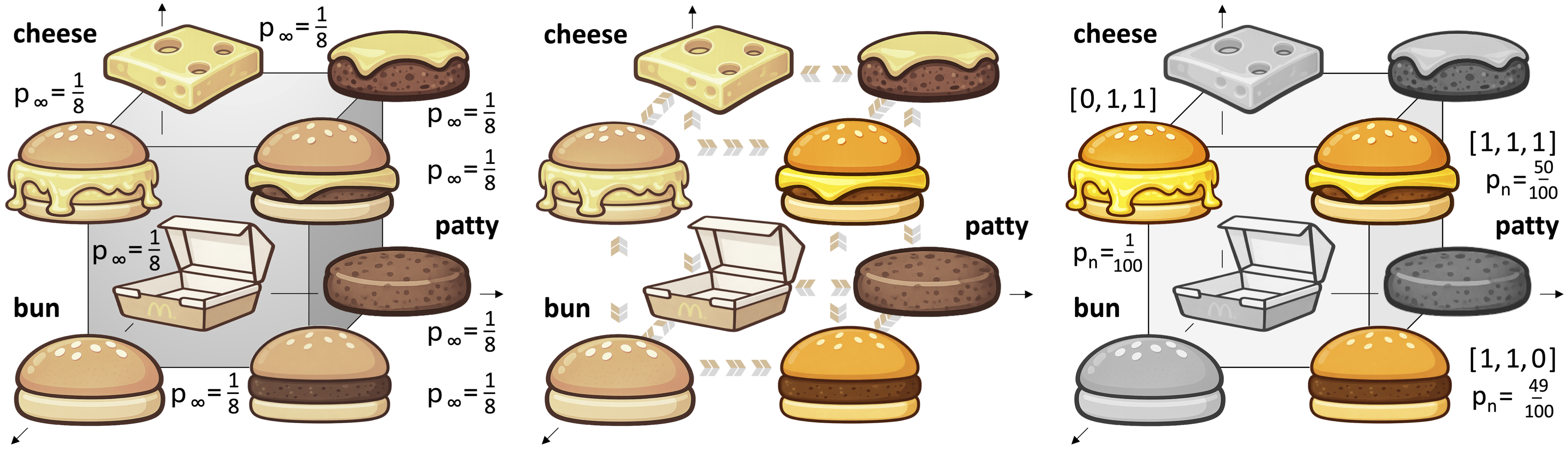} 
\caption{{\bf{\sffamily{Discrete modeling of reverse diffusion.}}} 
Noised data with maximum entropy and equal probabilities 
$p_{\infty} =\frac{1}{8}$
of all eight burgers (left);
reverse diffusion with probability-weighted flipping of the three ingredients gradually diffuses probabilities against their gradients 
towards the training set (middle);
de-noised data with three possible burgers,
cheeseburger    $\vec{x}_1 = [\,1,1,1\,]$ and
hamburger       $\vec{x}_2 = [\,1,1,0\,]$ and newly discovered
cheese sandwich $\vec{x}_3 = [\,0,1,1\,]$ 
with probabilities 
$p_{\text{data}}(\vec{x}_1) = 0.50$ and
$p_{\text{data}}(\vec{x}_2) = 0.49$ and
$p_{\text{data}}(\vec{x}_3) = 0.01$,
highlighted in color (right).}
%Learned mask.
\label{fig03}
\end{figure}

%%%%%%%%%%%%%%%%%%%%%%%%%%%%%%%%%%%%%%%%%%%%%%%%%%%%%%%%%%%%%%%%%%%%%%%%%%
\noindent
While the forward diffusion process {\it{destroys structure}}
by randomly flipping ingredients,
the reverse process aims to {\it{reconstruct structure}}
from partially corrupted burgers.
The reverse process is governed by the time-reversed transition probability
$p(\vec{x}_{t-1}\mid \vec{x}_t)$,
the probability that the previous burger at time 
$(t-1)$ was $\vec{x}_{t-1}\in\{0,1\}^3$ 
given the observed burger 
at time $t$ is $\vec{x}_t\in\{0,1\}^3$.
For the three-ingredient burger problem, 
we can calculate 
the reverse kernel analytically using {\it{Bayes' theorem}},
\begin{equation}
   p(\vec{x}_{t-1} \,|\, \vec{x}_t)
= \frac{p(\vec{x}_t \,|\, \vec{x}_{t-1}) \, p_{t-1}(\vec{x}_{t-1})}
 {\sum_{\vecs{x}} p(\vec{x}_t \,|\, \vec{x})\,p_{t-1}(\vec{x})},
\label{reverse_discrete}
\end{equation}
where 
%%%
$p(\vec{x}_{t-1} \,|\, \vec{x}_t)$
is the {\it{posterior}}, 
the probability of the previous state $\vec{x}_{t-1}$
given the current state $\vec{x}_t$;
%%%
$p(\vec{x}_t\mid \vec{x}_{t-1})$ 
is the {\it{likelihood}},
the forward transition probability from equation (\ref{discrete_one_step});
%%%
$p_{t-1}(\vec{x}_{t-1})$ 
is the {\it{prior}},
the distribution of states at time $(t-1)$
obtained from the forward solution 
in equations (\ref{evolution_p}) and (\ref{evolution_hamming}); and 
%%%
${\sum_{\vecs{x}} p(\vec{x}_t \,|\, \vec{x})\,p_{t-1}(\vec{x})}$
is the {\it{evidence}},
the sum over all 
possible burger configurations $\vec{x} \in \{0,1\}^3$
that normalizes the probabilities to sum up to one. 
For the three-ingredient benchmark, 
we can evaluate the reverse kernel {\it{exactly}}, 
because the state space contains only $2^3=8$ configurations. 
For the full 146-ingredient problem in Section \ref{sec04}, 
this exact computation becomes intractable,
because the corresponding normalization sum
$\sum_{\vecs{x}}$ 
would run over $2^{146}$ possible configurations.
In this case, we will evaluate the reverse kernel {\it{approximately}}
by learning a model 
$p_{\vecs{\theta}}(\vec{x}_{t-1} \,|\, \vec{x}_t)$,
for example a neural network with trainable parameters~$\vec{\theta}$.
In intuitive terms,
if the forward process gradually moves probability mass
away from the two training burgers,
the reverse process learns how to move probability mass back toward them
(Fig.~\ref{fig03}).\\[6.pt]
%%%%%%%%%%%%%%%%%%%%%%%%%%%%%%%%%%%%%%%%%%%%%%%%%%%%%%%%%%%%%%%%%%%%%%%%%%
{\textsf{\textbf{Numerical reverse sampling.}}}
%%%%%%%%%%%%%%%%%%%%%%%%%%%%%%%%%%%%%%%%%%%%%%%%%%%%%%%%%%%%%%%%%%%%%%%%%%
To generate trajectories from the reverse process, 
we sample sequentially from the conditional distributions 
$p (\vec{x}_{t-1} \mid \vec{x}_t)$ in equation (\ref{reverse_discrete}).
Starting from a noisy configuration $\vec{x}_T$
that we typically draw from an approximately uniform distribution, 
we iteratively generate a sequence
$\vec{x}_T \to \vec{x}_{T-1} \to \cdots \to \vec{x}_0$,
by sampling 
$\vec{x}_{t-1}$ from 
$p(\vec{x}_{t-1} \mid \vec{x}_t)$ 
at each step. 
This procedure defines a stochastic trajectory that progressively reconstructs structure by transporting probability mass from high-entropy states toward the low-entropy data distribution.
%%%%%%%%%%%%%%%%%%%%%%%%%%%%%%%%%%%%%%%%%%%%%%%%%%%%%%%%%%%%%%%%%%%%%%%%
 \\[6.pt]
%%%%%%%%%%%%%%%%%%%%%%%%%%%%%%%%%%%%%%%%%%%%%%%%%%%%%%%%%%%%%%%%%%%%%%%%
%%%%%%%%%%%%%%%%%%%%%%%%%%%%%%%%%%%%%%%%%%%%%%%%%%%%%%%%%%%%%%%%%%%%%%%%%%%
%{\it{\textsf{\textbf{Burger interpretation.}}}
%%%%%%%%%%%%%%%%%%%%%%%%%%%%%%%%%%%%%%%%%%%%%%%%%%%%%%%%%%%%%%%%%%%%%%%%%%%
%In the continuous burger example,
%the true reverse distribution reflects the structure
%of the training data.
%Recall that under $p_{\mathrm{data}}(\vec{x})$, 
%${\textsf{P}}(x_\mathrm{bun}=1)  =1$ and
%${\textsf{P}}(x_\mathrm{patty}=1)=1$ and
%${\textsf{P}}(x_\mathrm{cheese}=1)= \mbox{$\frac{1}{2}$}$.
%Suppose we observe a noisy burger,
%$\vec{x}_t = [\,1,0,1\,]$, a cheese sandwich. 
%Under the true reverse distribution,
%it is very unlikely that the previous burger
%had no patty, $x_{\mathrm{patty}}=0$, 
%because the patty was always present in the data, $x_{\mathrm{patty}}=1$
%Therefore,
%$q(x_{{\mathrm{patty}}, t-1}=1 \mid \vec{x}_t) \approx 1$.
%If the model predicts a smaller value,
%the cross-entropy loss increases.
%Training pushes the model to align
%with the data-induced structure.
%%
%Thus, minimizing the cross-entropy loss 
%ensures that the reverse process 
%restores the deterministic ingredients, bun and patty,
%while allowing stochastic variation in cheese.}\\[6.pt]
%%%%%%%%%%%%%%%%%%%%%%%%%%%%%%%%%%%%%%%%%%%%%%%%%%%%%%%%%%%%%%%%%%%%%%%%%%
{\it{\textsf{\textbf{Geometric interpretation.}}}
%%%%%%%%%%%%%%%%%%%%%%%%%%%%%%%%%%%%%%%%%%%%%%%%%%%%%%%%%%%%%%%%%%%%%%%%%%
The forward process defines
a {\it{random walk}} on the cube:
Transitions along the edges are symmetric,
all directions are equally likely, and
probability spreads uniformly across the cube.
The reverse process defines
a {\it{biased random walk}} on the cube:
Transitions are no longer symmetric,
but depend on the data distribution.
Edges that point toward high-density vertices,
$[\,1,1,0\,]$ and $[\,1,1,1\,]$,
have a higher transition probability,
while edges pointing away from the data manifold
have a lower transition probability.
Bayes' theorem 
assigns transition probabilities 
along the edges of the reverse random walk 
so that probability 
mass flows preferentially toward the training vertices, 
$[\,1,1,0\,]$ and $[\,1,1,1\,]$.
In the long-time limit,
this biased random walk concentrates probability
near the initial data support.
It reduces entropy and restores structure (Fig. \ref{fig03}).}
%%%%%%%%%%%%%%%%%%%%%%%%%%%%%%%%%%%%%%%%%%%%%%%%%%%%%%%%%%%%%%%%%%%%%%%%%%
\subsection{\textsf{\textbf{Discrete forward and reverse diffusion}}}
\begin{figure}[h]
\centering
\includegraphics[width=1.00\textwidth]{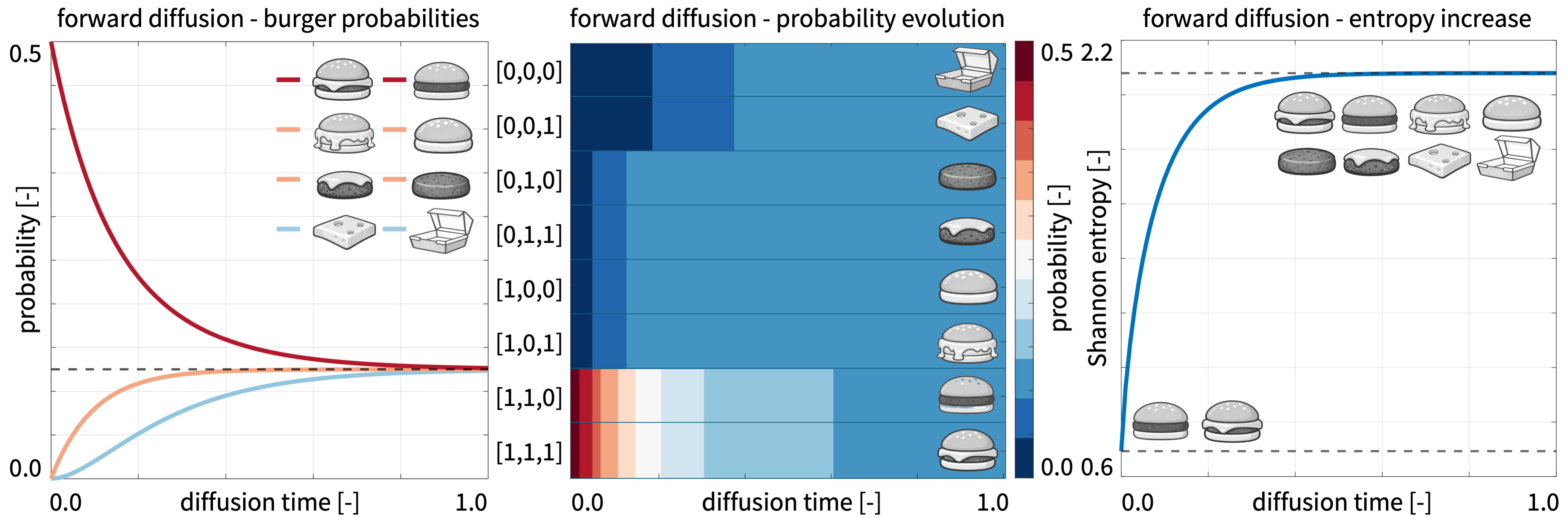} 
\caption{{\bf{\sffamily{Forward diffusion in discrete ingredient space.}}} 
Forward diffusion noises the training data to obtain a uniform distribution. 
Starting from two training states, cheeseburger and hamburger, probability mass spreads across all states and approaches a uniform distribution 
as reflected by the
convergence of individual state probabilities (left), 
homogenization of the probability distribution (center), and 
increase in Shannon entropy (right).}
\label{fig04}
%\end{figure} %\\[6.pt]
%%%%%%%%%%%%%%%%%%%%%%%%%%%%%%%%%%%%%%%%%%%%%%%%%%%%%%%%%%%%%%%%%%%%%%%%
\vspace*{0.4cm}
%%%%%%%%%%%%%%%%%%%%%%%%%%%%%%%%%%%%%%%%%%%%%%%%%%%%%%%%%%%%%%%%%%%%%%%%
%\begin{figure}[h]
\centering
\includegraphics[width=1.00\textwidth]{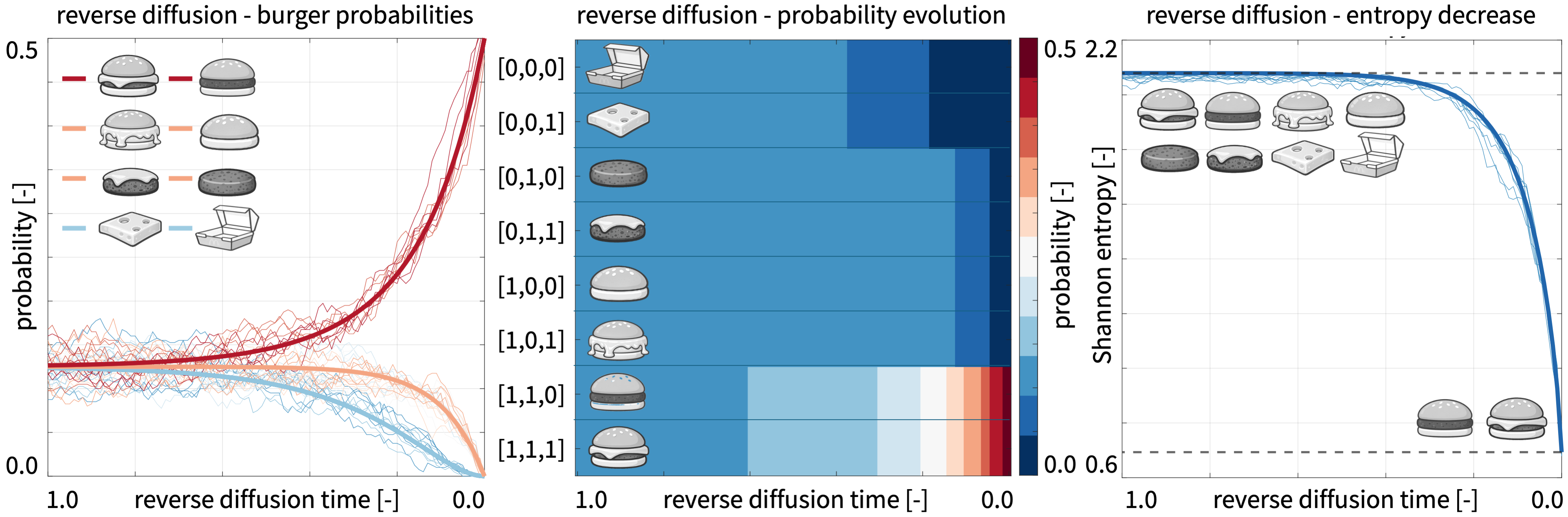} 
\caption{{\bf{\sffamily{Reverse diffusion in discrete ingredient space.}}} 
Reverse diffusion recovers the training distribution from the uniform state. 
Starting from the noised state, probability mass concentrates back onto the original two burgers 
as reflected by the
divergence of state probabilities (left), 
re-localization in probability space (center), and 
rapid entropy decrease (right). 
Thin lines indicate stochastic realizations; 
thick lines show ensemble averages.}
\label{fig05}
\end{figure}

%%%%%%%%%%%%%%%%%%%%%%%%%%%%%%%%%%%%%%%%%%%%%%%%%%%%%%%%%%%%%%%%%%%%%%%%%%
\noindent
We now illustrate the behavior of discrete diffusion 
for the eight possible burgers defined by binary ingredient selection 
(Figs. \ref{fig04} and \ref{fig05}). 
The {\it{forward diffusion}} problem (\ref{discrete_one_step}) 
admits a closed-form solution: 
For an initial configuration $\vec{x}_0$, 
the distribution $p_t$ at time $t$ 
only depends on the Hamming distance $d(\vec{x}_0,\vec{x})$ 
and takes the explicit form
$ p_t(\vec{x} \mid \vec{x}_0)
= q_t^{\,d(\vecs{x}_0,\vecs{x})}
 [\, 1-q_t \,]^{\,3-d(\vecs{x}_0,\vecs{x})}$,
where
$q_t = \mbox{$\frac{1}{2}$}\, [\,1- [\,1 - 2\beta\,]^t \,]$ 
represents the cumulative flip probability after $t$ diffusion steps.
%For a general initial distribution 
%$p_0(\vec{x})$, the solution follows by linear superposition, 
%$p_t(\vec{x}) = \sum_{\vecs{x}_0} p_0(\vec{x}_0)\,p_t(\vec{x}\mid\vec{x}_0)$, and converges to the uniform distribution as $t \to \infty$.
For the {\it{reverse diffusion}} problem (\ref{reverse_discrete}), 
we generate trajectories Bayes' theorem
by sampling sequentially from the posterior transition probabilities 
$p(\vec{x}_{t-1} \mid \vec{x}_t)$, 
which we obtain from Bayes' theorem
using the forward transition probabilities and marginals,
where the dependence on the initial data $\vec{x}_0$ is implicitly encoded 
through the forward marginals $p_{t-1}(\vec{x})$.
This procedure produces stochastic paths 
$\vec{x}_T \to \vec{x}_{T-1} \to \cdots \to \vec{x}_0$ that reconstruct likely ingredient configurations consistent with the forward diffusion process. 
%%%%%%%%%%%%%%%%%%%%%%%%%%%%%%%%%%%%%%%%%%%%%%%%%%%%%%%%%%%%%%%%%%%%%%%%
 %\\[6.pt]
%%%%%%%%%%%%%%%%%%%%%%%%%%%%%%%%%%%%%%%%%%%%%%%%%%%%%%%%%%%%%%%%%%%%%%%%
During forward diffusion (Fig. \ref{fig04}), the model begins with a degenerate distribution concentrated on two equally probable training burgers, cheeseburger and hamburger, with $p_{0} = 0.500$ (red lines), and progressively spreads probability mass towards the cheese sandwich, plain bun, cheese patty, and plain patty (orange lines), and with a slight delay towards the plain cheese and the empty burger box (blue lines). Over time, the distribution approaches uniformity with $p_{\infty} = 0.125$, as evidenced by the convergence of all state probabilities toward the same value and the corresponding increase in Shannon entropy to $H(p_{\infty}) = 2.079$. This reflects the role of forward diffusion as a mixing process, which gradually removes information about the initial configuration.
During reverse diffusion (Fig. \ref{fig05}), the model inverts the diffusion dynamics. Starting from an approximately uniform distribution with $p_{\infty} = 0.125$, reverse diffusion progressively concentrates probability mass back onto the training states. This manifests itself in a separation of probabilities toward the cheeseburger and hamburger
that converge towards $p_{0} = 0.500$ as the entropy decreases towards $H(p_0) =0.6931$. The thin lines of the individual stochastic trajectories visualize the variability, while the thick lines of the ensemble average follow a smooth and consistent contraction toward the data distribution. Taken together, these results demonstrate that in the discrete setting, diffusion operates over a finite combinatorial state space, where all configurations are accessible during forward diffusion and the learned reverse dynamics selectively reconstruct the observed data modes.\\[6.pt]
%%%%%%%%%%%%%%%%%%%%%%%%%%%%%%%%%%%%%%%%%%%%%%%%%%%%%%%%%%%%%%%%%%%%%%%%%%
{\it{\textsf{\textbf{Analogy to mechanics.}}}
%%%%%%%%%%%%%%%%%%%%%%%%%%%%%%%%%%%%%%%%%%%%%%%%%%%%%%%%%%%%%%%%%%%%%%%%%%
The increase of entropy during forward diffusion 
reflects the loss of information and the approach to equilibrium, 
consistent with the second law of thermodynamics. 
Conversely, reverse diffusion reduces entropy by reintroducing structure 
through a learned drift term and 
effectively acts as a data-driven inverse process.
In this sense, 
it resembles transport against concentration gradients, $\nabla p$,
or {\it{uphill diffusion}}, 
where probability mass flows toward higher-density regions 
under the action of an externally learned driving force. 
This is analogous to {\it{driven transport processes}} in mechanics,
such as unmixing processes that restore structure 
from a mixed state under an external force.}%\\[6.pt]
%%%%%%%%%%%%%%%%%%%%%%%%%%%%%%%%%%%%%%%%%%%%%%%%%%%%%%%%%%%%%%%%%%%%%%%%%%
\subsection{\textsf{\textbf{Generating new burgers by discrete diffusion}}}
\begin{figure}[h]
\centering
\includegraphics[width=1.00\textwidth]{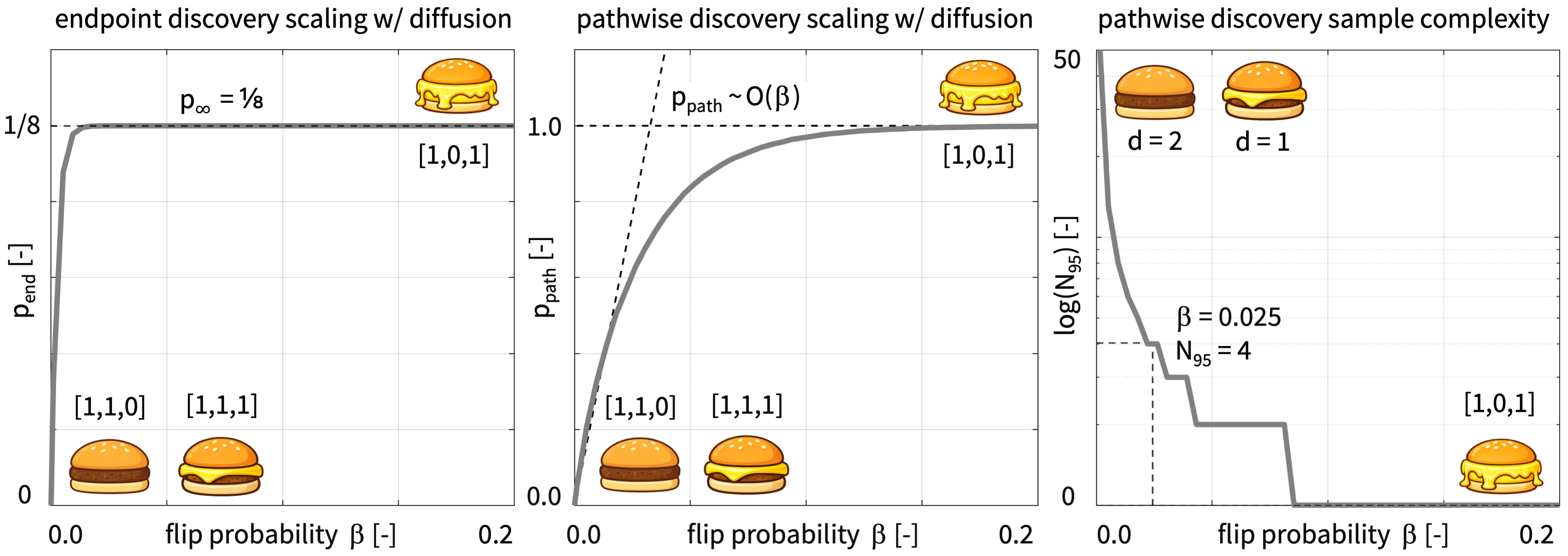} 
\caption{{\textsf{\textbf{Generating new burgers by discrete diffusion.}}} 
Sampling complexity for discovering a new burger, a cheese sandwich 
$[\,1,0,1\,]$, with Hamming distances of $d$ = 2 and $d$ = 1 from 
the training data, hamburger $[\,1,1,0\,]$ and cheeseburger $[\,1,1,1\,]$.
Probability of endpoint discovery $p_{\text{end}}$ increases 
with flip probability $\beta$
and saturates at the uniform limit $p_{\infty} = \frac{1}{8}$ (left);
probability of pathwise discovery $p_{\text{path}}$ 
increases 
from an initial small-$\beta$ scaling, $p_{\text{path}} \approx {\mathcal{O}}(\beta)$,
and approach unity as trajectories explore the full state space (middle);
number of sample trajectories required 
for 95\% discovery $N_{95}$ 
decreases rapidly with increasing flip probability $\beta$ (right).}
\label{fig06}
\end{figure}

%%%%%%%%%%%%%%%%%%%%%%%%%%%%%%%%%%%%%%%%%%%%%%%%%%%%%%%%%%%%%%%%%%%%%%%%%%
\noindent 
We next quantify the ability of discrete diffusion to generate burgers not present in the training set. As a representative example, we consider the cheese sandwich $[1,0,1]$, which differs from the training burgers by one ingredient flip from the cheeseburger, $d$ = 1, and two flips from the hamburger, $d$ = 2.
We report sampling complexity 
via the probabilities $p_{\text{path}}$ and $p_{\text{end}}$ that 
quantify the likelihood that a single trajectory passes through or ends at the defined target burger, and  
via the number of independent samples required
to achieve 95\% probability of discovery 
$N_{95} = [\, {\rm{log}}(0.05)/{\rm{log}}(1-p)]$ 
after $T$ = 100 diffusion steps (Fig. \ref{fig06}).
%%%%%%%%%%%%%%%%%%%%%%%%%%%%%%%%%%%%%%%%%%%%%%%%%%%%%%%%%%%%%%%%%%%%%%%%
 %\\[6.pt]
%%%%%%%%%%%%%%%%%%%%%%%%%%%%%%%%%%%%%%%%%%%%%%%%%%%%%%%%%%%%%%%%%%%%%%%%
Discovery probability increases rapidly with increasing flip probability $\beta$ :
For small $\beta$, discovery is limited by the probability of rare multi-bit flips, while for large $\beta$, the process approaches uniform sampling over the state space.
As a results, the endpoint probability $p_{\rm end}$ increases with flip probability $\beta$ and approaches the uniform value $p_{\infty} = \frac{1}{8}$ that reflects complete mixing over the eight possible burgers. 
This discrete solution agrees with the analytical solution,
$p_{\rm end} = \mbox{$\frac{1}{2}$} \, q \, [1-q]$ with 
$q = \mbox{$\frac{1}{2}$} \, [1-[1-2\,\beta]^T]$,
where $T$ is the number of diffusion steps. 
In contrast, the pathwise probability $p_{\rm path}$ 
initially obeys small-$\beta$ scaling, $p_{\text{path}} \approx {\mathcal{O}}(\beta)$,
but then rises to approaches unity, since the cheese sandwich can be encountered at any intermediate step along a trajectory.
The corresponding discovery cost $N_{95}$ decreases sharply with increasing flip probability $\beta$. 
%%%%%%%%%%%%%%%%%%%%%%%%%%%%%%%%%%%%%%%%%%%%%%%%%%%%%%%%%%%%%%%%%%%%%%%%%%
\section{{\textsf{\textbf{Continuous diffusion}}}}\label{sec03}
%%%%%%%%%%%%%%%%%%%%%%%%%%%%%%%%%%%%%%%%%%%%%%%%%%%%%%%%%%%%%%%%%%%%%%%%%%
\noindent
We now move from discrete diffusion for ingredient selection
to {\it{continuous diffusion}} for {\it{ingredient quantification}}.
Instead of binary ingredients,
we consider real-valued ingredient weights,
\beq
\vec{w}^* = [\, w^*_{\text{bun}}, w^*_{\text{patty}}, w^*_{\text{cheese}} \,]
\; \in \; \mathbb{R}^3,
\eeq
where each component denotes the weight in grams
of the ingredient.
For convenience, we normalize all weights by the weights of 
a normal sized bun $w^0_{\text{bun}}=55\,\text{g}$, 
a small patty $w^0_{\text{patty}}=45\,\text{g}$, and 
a slice of cheese $w^0_{\text{cheese}}=14\,\text{g}$, 
and translate the burger into a coordinate system with the cheeseburger at the origin, 
\beq
\vec{w} 
= [\, w_{\text{bun}}, w_{\text{patty}}, w_{\text{cheese}} \,]
= [\, 
w^*_{\text{bun}}/w^0_{\text{bun}}, 
w^*_{\text{patty}}/w^0_{\text{patty}}, 
w^*_{\text{cheese}}/w^0_{\text{cheese}} \,] - [\,1,1,1\,]
\; \in \; \mathbb{R}^3 \,.
\eeq
In these two coordinate systems, 
the cheeseburger is represented as
$\vec{w}_1^* = [\, 55\,\text{g}, 45\,\text{g}, 14\,\text{g} \,]$ and 
$\vec{w}_1 = [\,0,0,0\,]$
and the hamburger is 
$\vec{w}_2^* = [\, 55\,\text{g}, 45\,\text{g}, 0\,\text{g} \,]$ and 
$\vec{w}_2 = [\,0,0,-1\,]$.\\[6.pt]
%%%%%%%%%%%%%%%%%%%%%%%%%%%%%%%%%%%%%%%%%%%%%%%%%%%%%%%%%%%%%%%%%%%%%%%%%%
{\textsf{\textbf{Training data.}}}
%%%%%%%%%%%%%%%%%%%%%%%%%%%%%%%%%%%%%%%%%%%%%%%%%%%%%%%%%%%%%%%%%%%%%%%%%%
The training data consist of $N$ burgers,
$\{ \vec{w}_i \}_{i=1}^N$,
each representing a set of ingredient weights.
For this case, the empirical data distribution is
$p_{\text{data}}(\vec{w})
=
\mbox{$\frac{1}{N} \sum_{i=1}^N$}
\delta(\,\vec{w} - \vec{w}_i\,)$,
where, in the continuous case, 
$\delta$ is the Dirac delta distribution.
Here, similar to the discrete case, we assume that the training data consist of two burgers,
\beq
\vec{w}_1 = [\,0,0,0\,] 
\quad \mbox{and} \quad
\vec{w}_2 = [\,0,0,-1] \,,
\eeq
a cheeseburger with a standard size bun, patty, and cheese, and 
a hamburger with a standard size bun and patty, but no cheese, 
with equal probabilities,
\beq
p_{\text{data}}(\vec{w})
= \mbox{$\frac{1}{2}$} \delta(\,\vec{w} - \vec{w}_1\,)
+ \mbox{$\frac{1}{2}$} \delta(\,\vec{w} - \vec{w}_2\,) \,,
\eeq
such that 
$p_{\text{data}}(\vec{w}_1) = \frac{1}{2}$ and
$p_{\text{data}}(\vec{w}_2) = \frac{1}{2}$.\\[6.pt]
%%%%%%%%%%%%%%%%%%%%%%%%%%%%%%%%%%%%%%%%%%%%%%%%%%%%%%%%%%%%%%%%%%%%%%%%%%
{\it{\textsf{\textbf{Geometric interpretation.}}}
%%%%%%%%%%%%%%%%%%%%%%%%%%%%%%%%%%%%%%%%%%%%%%%%%%%%%%%%%%%%%%%%%%%%%%%%%%
The three continuous ingredient weights define a three-dimensional state
space $\mathbb{R}^3$, with an infinite number of possible burgers, where the cheeseburger defines the origin, burgers with lower ingredient weights like the cheese-less hamburger have negative coordinates, and burgers with higher ingredient weights like the two-patty Mc Double have positive coordinates.
Unlike in the discrete case,
the state space is now continuous.
The discrete corners of the cube 
are replaced by the entire
three-dimensional Euclidean space.
Each burger is now a point in $\mathbb{R}^3$.
The training data form a low-dimensional cloud
within this space, 
in our example, represented by only two points.}
%%%%%%%%%%%%%%%%%%%%%%%%%%%%%%%%%%%%%%%%%%%%%%%%%%%%%%%%%%%%%%%%%%%%%%%%%%
\subsection{\textsf{\textbf{Forward diffusion}}} 
%%%%%%%%%%%%%%%%%%%%%%%%%%%%%%%%%%%%%%%%%%%%%%%%%%%%%%%%%%%%%%%%%%%%%%%%%%
\noindent
Forward diffusion gradually destroys structure 
by noising the training data. 
It modifies each burger point by random Gaussian noise, 
and at the same time, gently pulls it toward the origin. 
Specifically, we introduce continuous forward diffusion 
through a stochastic differential equation,
\beq
  \sca{d} \vec{w}_t
= \vec{f}(\vec{w}_t,t) \;\sca{d}t
+ g(t)\;\sca{d} \vec{B}_t,
\eeq
where
$\vec{w}_t \in \mathbb{R}^3$ is the ingredient-weight vector at diffusion time $t$,
with components 
$\vec{w}_t = [\,w_{\text{bun}}(t),\,w_{\text{patty}}(t),\,w_{\text{cheese}}(t)\,]$,
$\sca{d} \vec{w}_t$
is the {\it{stochastic differential}}
that represents the infinitesimal random increment of the process
that consists of a deterministic drift and a stochastic fluctuation,
$\vec{f}(\vec{w}_t,t)$ is the drift vector that defines the systematic trend,
$g(t)$ is the diffusion coefficient that defines the noise magnitude, and
$\sca{d}\vec{B}_t$ is the vector-valued stochastic differential of Brownian motion.
The {\it{Brownian motion}} of the continuous diffusion case
represents a continuous-time limit of 
the {\it{multinomial transitions}} or ingredient flipping
of the discrete diffusion case.\\[6.pt]
%%%%%%%%%%%%%%%%%%%%%%%%%%%%%%%%%%%%%%%%%%%%%%%%%%%%%%%%%%%%%%%%%%%%%%%%%%
{\textsf{\textbf{Ornstein--Uhlenbeck forward process.}}}
%%%%%%%%%%%%%%%%%%%%%%%%%%%%%%%%%%%%%%%%%%%%%%%%%%%%%%%%%%%%%%%%%%%%%%%%%%
Here we model diffusion through a variance-preserving {\it{Ornstein--Uhlenbeck process}} \cite{uhlenbeck1930},
because it provides a linear Gaussian diffusion with a closed-form solution that enables analytically tractable forward and reverse processes,
\beq
  \sca{d}\vec{w}_t
=-\mbox{$\frac{1}{2}$} \, \beta(t)\, \vec{w}_t \, \sca{d}t
+ \sqrt{\beta(t)}\, \sca{d} \vec{B}_t \,.
\label{OUsmall}
\eeq
The first term,
$-\tfrac{1}{2}\beta(t)\vec{w}_t$, 
is the {\it{drift}} 
that removes information 
by contracting all weights toward zero.
The second term,
$\sqrt{\beta(t)}\,d \vec{B}_t$, 
is the {\it{stochastic diffusion}} 
that increases entropy 
by injecting isotropic Gaussian noise
to spread the weights outward.
The parameter
$\beta(t)>0$ is the noise schedule or diffusion rate
that controls 
both the strength of fluctuations and the timescale of drift.
It increases as time progresses,
from an initial weak contraction and weak noise 
towards a strong contraction and strong noise. 
This stochastic differential equation 
has the following closed-form conditional distribution,
\beq
p( \vec{w}_t |\, \vec{w}_0)
=
\mathcal{N} 
(\, \vec{w}_t |\, \mu(t)\,\vec{w}_0 , \sigma(t)\vec{I} \,) \,,
\label{small_gauss}
\eeq
with
\beq
\mu(t) = \exp (\,- \mbox{$\frac{1}{2}$}{\alpha(t)} \,)
\quad \mbox{and} \quad
\sigma(t) = 1 - \exp(-\alpha(t)) 
\quad \mbox{and} \quad
\alpha(t)= \mbox{$\int_0^t$} \; \beta(s)\,\sca{d}s \,,
\eeq
where
$\mu(t)$ is the signal attenuation factor
that determines how much of $\vec{w}_0$ remains,
$\sigma(t)$ is the noise variance 
that is accumulated by time $t$, and
$\alpha(t)$ is the integrated noise.\\[6.pt]
%%%%%%%%%%%%%%%%%%%%%%%%%%%%%%%%%%%%%%%%%%%%%%%%%%%%%%%%%%%%%%%%%%%%%%%%%%
{\it{\textsf{\textbf{Analogy to mechanics.}}}
The Ornstein--Uhlenbeck process is equivalent to a linear Langevin equation that governs stochastic relaxation under competing dissipation and fluctuations. 
At the ensemble level, the process satisfies a Fokker--Planck equation that describes probability transport under linear drift and isotropic diffusion \cite{fokker1914,planck1917}.
Here, $\beta$ acts as a relaxation rate that controls both noise intensity and drift strength, and sets the timescale of mixing. 
Increasing $\beta$ results in faster exploration of the state space over finite time horizons, while leaving the equilibrium distribution unchanged.}
\\[6.pt]
%%%%%%%%%%%%%%%%%%%%%%%%%%%%%%%%%%%%%%%%%%%%%%%%%%%%%%%%%%%%%%%%%%%%%%%%%%
{\textsf{\textbf{One-step all-weight diffusion.}}}
%%%%%%%%%%%%%%%%%%%%%%%%%%%%%%%%%%%%%%%%%%%%%%%%%%%%%%%%%%%%%%%%%%%%%%%%%%
As illustrative example, 
we assume a three-ingredient burger, with initial ingredients
\beq
\vec{w}_0^* = [\, 55\,\text{g}, 45\,\text{g}, 0\,\text{g} \,]
\quad
\mbox{or}
\quad
\vec{w}_0 = [\, 0,0,-1 \,]\,,
\eeq
which corresponds to 
a hamburger with a 55\,g bun and a 45\,g patty. 
For Ornstein-Uhlenbeck forward diffusion, 
the noisy burger is
$\vec{w}_t = \mu(t)\,\vec{w}_0 + \sqrt{\sigma(t)}\, \vec{\epsilon}$
with
$\vec{\epsilon} \sim \mathcal{N}(0,I)$.
We choose a representative diffusion time such that
$\alpha(t) = \log (2)$,
and assume a Gaussian noise realization of
$\vec{\epsilon} = [\,0, \sqrt{2}, 1\,]$. 
For this example, 
the signal attenuation factor $\mu(t)$ and 
the noise variance $\sigma(t)$ become,
$ \mu(t) 
= 0.707$
and
$\sigma(t) 
= 0.5$.
The resulting mean,
$\mu(t)\,\vec{w}_0
= [\,0,0,-0.707 \,]$,
and injected noise,
$
\sqrt{\sigma(t)}\,\vec{\epsilon}
= [\,0.000, 1.000, 0.707 \,]$,
produce a noisy burger with the following weights,
\beq
\vec{w}_t
= [\,0, 1, 0\,]
\quad \mbox{or} \quad
\vec{w}_t^* 
= [\, 55\text{g},90\text{g},14\text{g} \,].
\eeq
The initial hamburger 
$\vec{w}_0 = [\, 0,0,-1 \,]$
turns into a Mc Double  
$\vec{w}_t = [\, 0,1,0 \,]$,
with a 55g bun and two 45g patties and 14g cheese. 
This diffusion corresponds to a simultaneous perturbation 
along the patty and cheese axis.
The signal attenuation factor $\mu$
shrinks the hamburger to the origin, the cheeseburger, 
in analogy to an exponential damping.
The noise variance $\sigma$
controls the magnitude of the injected noise 
and grows monotonically with time.
Here we applied a large noise realization of
$\vec{\epsilon}=[\,0, \sqrt{2}, 1\,]$ in a single step. 
Typically $\vec{\epsilon}$ is much smaller,
and it would be highly unlikely to reach the Mc Double
within a small number of steps.  \\[6.pt]
%%%%%%%%%%%%%%%%%%%%%%%%%%%%%%%%%%%%%%%%%%%%%%%%%%%%%%%%%%%%%%%%%%%%%%%%%%
{\it{\textsf{\textbf{Geometric interpretation}}}
%%%%%%%%%%%%%%%%%%%%%%%%%%%%%%%%%%%%%%%%%%%%%%%%%%%%%%%%%%%%%%%%%%%%%%%%%%
Geometrically, each burger is a point in $\mathbb{R}^3$,
with coordinates given by its ingredient weights.
Forward diffusion maps each point to a Gaussian cloud centered at $\mu(t) \vec{w}_0$.
As the mean $\mu(t)$ decreases,
the center contracts toward the origin.
As the variance $\sigma(t)$ increases,
the cloud expands towards a sphere.
As time grows, 
distinct burgers become less distinguishable, 
because their Gaussian clouds overlap more strongly, 
and the distribution smoothens out.
In the long-time limit, 
the influence of the initial data vanishes 
and the distribution approaches 
an isotropic spherical Gaussian distribution
centered at the origin, 
that represents maximum uncertainty about the ingredient weights.
This is the continuous analogue of the discrete random walk on the cube:
Instead of 
probability mass spreading discretely between vertices, 
probability density spreads continuously in $\mathbb{R}^3$ and 
loses the sharp structure of the training data.}\\[6.pt]
%%%%%%%%%%%%%%%%%%%%%%%%%%%%%%%%%%%%%%%%%%%%%%%%%%%%%%%%%%%%%%%%%%%%%%%%
{\it{\textsf{\textbf{Analogy to mechanics.}}} 
%%%%%%%%%%%%%%%%%%%%%%%%%%%%%%%%%%%%%%%%%%%%%%%%%%%%%%%%%%%%%%%%%%%%%%%%%%
We can describe 
the forward diffusion process 
by a Fokker--Planck equation,
${\partial p}/{\partial t}
= -\nabla \cdot (\vec{b}\, p)
+ \frac{1}{2}\nabla \cdot (\, \ten{D} \cdot \nabla p \,)$,
where 
$p(\vec{w},t)$ is  the probability density of burgers with ingredient weights \vec{w} at time $t$,
$\vec{b} = -\tfrac{1}{2} \, \beta \, \vec{w}$ is the drift, and
$\ten{D} = \beta \, \ten{I}$ is the diffusion tensor. 
For a constant diffusion tensor $\ten{D} =\beta \, \ten{I}$, 
this reduces to a drift--diffusion equation,
${\partial p}/{\partial t}
= \frac{1}{2} \, \beta \, \nabla^2 p 
- \nabla \cdot (\,\tfrac{1}{2}\beta \, \vec{w} \, p \, )$,
that highlights the combined effects 
of isotropic diffusion and a linear restoring force.
This structure is directly analogous to transport equations in continuum mechanics, 
where advection and diffusion jointly govern the evolution of conserved quantities.
Continuous forward diffusion takes the interpretation 
of the continuum limit of the discrete Markov process, 
in which the generator of the random walk converges to a diffusion operator. 
In this sense, forward diffusion provides a bridge 
between stochastic hopping on a finite state space 
and transport in a continuous ingredient-weight space, 
analogous to the transition from lattice models 
to continuum models in classical mechanics.}
%%%%%%%%%%%%%%%%%%%%%%%%%%%%%%%%%%%%%%%%%%%%%%%%%%%%%%%%%%%%%%%%%%%%%%%%%%
\subsection{\textsf{\textbf{Reverse diffusion}}}
%%%%%%%%%%%%%%%%%%%%%%%%%%%%%%%%%%%%%%%%%%%%%%%%%%%%%%%%%%%%%%%%%%%%%%%%%%
\noindent
While forward diffusion gradually destroys structure 
by repeatedly noising the training data, 
reverse diffusion aims to reconstruct structure
by progressively sharpening noisy samples.
% XXXX
Instead of pushing burger points toward the origin and injecting noise, 
the reverse process 
defines a drift,
which in general must be learned from data,
that pulls noisy weight vectors 
back toward regions of high data density.
Specifically, we introduce 
continuous reverse diffusion 
through a time-reversed stochastic differential equation 
that incorporates a {\it{score function}},
\beq
  \sca{d} \vec{w}_t
= [\, g^2(t) \nabla_{\vecs{w}} \log (p_t(\vec{w}_t))
- \vec{f}(\vec{w}_t, t) \, ] \sca{d}t
+ g(t) \sca{d} \tilde{\vec{B}}_t \, ,
\eeq
where
$\nabla_{\vecs{w}} \log (p_t(\vec{w}))$ is the score function
in terms of $p_t(\vec{w})$, the marginal distribution of $\vec{w}_t$, and
$\sca{d}\tilde{\vec{B}}_t$ is reverse-time vector-valued stochastic differential of Brownian motion.\\[6.pt]
%%%%%%%%%%%%%%%%%%%%%%%%%%%%%%%%%%%%%%%%%%%%%%%%%%%%%%%%%%%%%%%%%%%%%%%%%%
{\textsf{\textbf{Ornstein–Uhlenbeck reverse process.}}}
%%%%%%%%%%%%%%%%%%%%%%%%%%%%%%%%%%%%%%%%%%%%%%%%%%%%%%%%%%%%%%%%%%%%%%%%%%
Here, in analogy with the forward diffusion process, 
for reverse diffusion,
we use the {\it{time-reverse Ornstein-Uhlenbeck process}} 
\cite{anderson1982},
\beq
 \sca{d} \vec{w}_t
=[\,
 \mbox{$\frac{1}{2}$} \, \beta(t) \, \vec{w}_t
+\beta(t) \nabla_{\vecs{w}} \log (p_t(\vec{w}_t))
\,] \sca{d}t
+ \sqrt{\beta(t)}\, \sca{d} \tilde{\vec{B}}_t.
\label{OUreverse_small}
\eeq
The first term, 
$\frac{1}{2}\beta(t)\vec{w}_t$,
is the {\it{reverse drift}}
that counteracts the contraction of the forward process 
and re-expands the state away from the origin. 
The second term,
$\beta(t)\nabla_{\vecs{w}} \log p_t(\vec{w}_t)$,
is the {\it{score term}}
that steers samples along the direction of increasing likelihood 
toward the data manifold.
Together, these two deterministic components define an effective probability force that reconstructs structure from noise. 
The third term, 
$\sqrt{\beta(t)}\,\mathrm{d}\tilde{\vec{B}}_t$,
is the {\it{stochastic diffusion}} 
that injects controlled randomness
to ensure sufficient exploration of the state space 
and prevent collapse onto a single trajectory. 
We can interpret 
reverse diffusion as a guided stochastic relaxation process 
that converts Gaussian noise into structured samples 
by following the gradient flow of the log-density 
while maintaining thermal fluctuations. \\[6.pt]
%%%%%%%%%%%%%%%%%%%%%%%%%%%%%%%%%%%%%%%%%%%%%%%%%%%%%%%%%%%%%%%%%%%%%%%%%%
{\textsf{\textbf{Score function.}}}
%%%%%%%%%%%%%%%%%%%%%%%%%%%%%%%%%%%%%%%%%%%%%%%%%%%%%%%%%%%%%%%%%%%%%%%%%%
Our example permits an analytic evaluation 
of the score function, $\nabla_{\vecs{w}} \log (p_t(\vec{w}))$,
because the linear Gaussian forward process 
transforms the discrete training distribution 
into a tractable Gaussian mixture, 
for which the log-density gradient admits a closed-form expression.
For a training set of \(N\) samples
\(
\{\vec{w}_i\}_{i=1}^N
\),
the empirical data distribution is
\(
p_{\rm data}(\vec{w})
=
\frac{1}{N}\sum_{i=1}^N \delta(\vec{w}-\vec{w}_i).
\)
Under forward Ornstein--Uhlenbeck diffusion, 
each point mass evolves into a Gaussian with mean $\mu(t)\vec{w}_i$
and covariance $\sigma(t)\ten{I}$,
and the forward marginal distribution at time $t$ 
is available in closed form,
\beq
p_t(\vec{w})
=
\mbox{$\frac{1}{N} \, \sum_{i=1}^N$} \,
\mathcal{N}\!\left(
\vec{w};
\mu(t)\vec{w}_i,
\sigma(t)\ten{I}
\right) ,
\eeq
with Ornstein--Uhlenbeck moments
$\mu(t)=\exp({-\frac{1}{2} \, \alpha(t)})$
and
$\sigma(t)=1-\exp({-\alpha(t)})$.
For the present example, 
the training set contains two burgers only, 
the cheeseburger with $\vec{w}_1=[\,0,0,0\,]$ and
the hamburger    with $\vec{w}_2=[\,0,0,-1\,]$,
and this expression reduces to
$
p_t(\vec{w})
=
\frac{1}{2}\,
\mathcal{N}\!\left(
\vec{w};
\mu(t)\vec{w}_1,
\sigma(t)\ten{I}
\right)
+
\frac{1}{2}\,
\mathcal{N}\!\left(
\vec{w};
\mu(t)\vec{w}_2,
\sigma(t)\ten{I}
\right)
$.
Taking the gradient of the log-density yields
\beq
  \nabla_{\vecs{w}} \log (p_t(\vec{w}))
= \sum_{i=1}^N
  \gamma_i(\vec{w},t) \,
  \frac{\mu(t)\vec{w}_i - \vec{w}}{\sigma(t)}
  \quad \mbox{with} \quad
  \gamma_i(\vec{w},t)
= \frac{
  \exp\!\left(
- \|\vec{w}-\mu(t)\vec{w}_i\|^2 / (2\sigma(t))
  \right)
  }{
  \sum_{j=1}^N
  \exp\!\left(
- \|\vec{w}-\mu(t)\vec{w}_j\|^2 / (2\sigma(t))
  \right) } \,,
\eeq
where $\gamma_i(\vec{w},t)$ 
denotes the posterior probability weight of component $i$,
that is, the probability that $\vec{w}$ originated
from the $i$-th training sample under the forward process.
This expression shows that the score is a weighted average of linear
restoring directions pointing from $\vec{w}$ toward the
forward-diffused training samples $\mu(t)\vec{w}_i$.
The sum in the denominator runs over the training data $N$.
For the three-ingredient benchmark, 
we can evaluate the score function {\it{exactly}},
because the discrete training data has only two configurations, 
cheeseburger and hamburger.
For the full 146-ingredient problem in Section \ref{sec04},
this exact computation becomes intractable, 
because the corresponding normalization sum would run over 2,260 training configurations.
In this case, we will evaluate the score function {\it{approximately}}
by learning a model $\vec{s}_{\vecs{\theta}}(\vec{w},t)$,
for example a neural network with trainable parameters $\vec{\theta}$.\\[6.pt]
%%%%%%%%%%%%%%%%%%%%%%%%%%%%%%%%%%%%%%%%%%%%%%%%%%%%%%%%%%%%%%%%%%%%%%%%%%
{\textsf{\textbf{Numerical reverse sampling.}}}
%%%%%%%%%%%%%%%%%%%%%%%%%%%%%%%%%%%%%%%%%%%%%%%%%%%%%%%%%%%%%%%%%%%%%%%%%%
To generate reverse trajectories, 
we discretize the reverse process with the {\it{Euler--Maruyama}} method
\cite{maruyama1954}.
At each reverse step $k$, we evaluate the score at the current state and at the corresponding forward time
\(
t = T-k\Delta t
\),
which decreases from \(T\) to \(0\) during sampling.
The numerical update is
\beq
\vec{w}_{k+1}
=
  \vec{w}_k
+ [\, \mbox{$\frac{1}{2}$} \, \beta \vec{w}_k
  + \beta \, \nabla_{\vecs{w}} \log (p_t(\vec{w}_k)) \, ]
  \Delta t
+ \sqrt{\beta \Delta t}\,\vec{\xi}_k
  \quad \mbox{with} \quad 
  \vec{\xi}_k \sim \mathcal{N}(\vec{0},\ten{I})
\eeq
where $\vec{\xi}_k$ is a standard Gaussian random vector.
This procedure transports samples 
from the noisy terminal mixture $p_T(\vec{w})$ 
back toward the empirical two-burger distribution at $t=0$.
This continuous reverse process is the analogue 
the discrete reverse process in Section \ref{sec02},
where we sample from the discrete posterior transition probabilities, 
for which Bayes’ rule defines the reverse kernel exactly.\\[6.pt]
%%%%%%%%%%%%%%%%%%%%%%%%%%%%%%%%%%%%%%%%%%%%%%%%%%%%%%%%%%%%%%%%%%%%%%%%%%
{\it{\textsf{\textbf{Analogy to mechanics.}}}
%%%%%%%%%%%%%%%%%%%%%%%%%%%%%%%%%%%%%%%%%%%%%%%%%%%%%%%%%%%%%%%%%%%%%%%%%%
We can interpret the reverse diffusion process 
as a stochastic gradient flow 
that reconstructs the data distribution, 
analogous to drift toward energy minima in dissipative mechanical systems. 
In the deterministic limit, this reduces to a gradient flow,
$\sca{d}{\vec{w}}/\sca{d}t = -\nabla U(\vec{w},t)$,
with $U(\vec{w},t) = -\log (p_t(\vec{w}))$, 
consistent with variational formulations of dissipative systems.
The learned score function plays the role of an effective driving force 
that guides trajectories toward regions of high probability density. 
In this interpretation, the dynamics take the form,
$\sca{d} \vec{w} 
= - \nabla U (\vec{w})\, \sca{d}t 
+ \sqrt{\beta}\, \sca{d} \vec{w}_t$,
with an effective potential, 
$U(\vec{w}) = -\log (p(\vec{w}))$, 
so that the score function, 
$\nabla \log (p(\vec{w}))$,
corresponds to a conservative force driving the system toward equilibrium,
while the additional linear term reflects the restoring drift inherited from the forward Ornstein--Uhlenbeck process.}\\[6.pt]
%%%%%%%%%%%%%%%%%%%%%%%%%%%%%%%%%%%%%%%%%%%%%%%%%%%%%%%%%%%%%%%%%%%%%%%%%%
{\textsf{\textbf{Relation to discrete diffusion.}}}
%%%%%%%%%%%%%%%%%%%%%%%%%%%%%%%%%%%%%%%%%%%%%%%%%%%%%%%%%%%%%%%%%%%%%%%%%%
In the discrete diffusion model,
the reverse process learns
a conditional probability,
$p_{\vecs{\theta}}
(\vec{x}_{t-1} \,|\, \vec{x}_t)$,
by minimizing a cross-entropy loss.
This loss is equivalent to minimizing
the Kullback--Leibler divergence between
the true and learned reverse transition kernels.
In the continuous diffusion model,
the reverse process learns
the score function,
$\nabla_{\vecs{w}} \log p_t(\vec{w}))$,
by minimizing a score-matching loss.
This loss is equivalent to minimizing
a time-integrated Kullback--Leibler divergence
between the true reverse-time stochastic process
and the model-implied reverse-time process.
In both cases,
forward diffusion increases entropy,
reverse diffusion learns dynamics that reduce entropy, and
training minimizes a divergence
between true and learned reverse processes.
The discrete cross-entropy loss
and the continuous score-matching loss
are two manifestations of the same principle:
learning the time-reversed dynamics of diffusion.\\[6.pt]
%%%%%%%%%%%%%%%%%%%%%%%%%%%%%%%%%%%%%%%%%%%%%%%%%%%%%%%%%%%%%%%%%%%%%%%%%%
{\it{\textsf{\textbf{Burger interpretation.}}}
%%%%%%%%%%%%%%%%%%%%%%%%%%%%%%%%%%%%%%%%%%%%%%%%%%%%%%%%%%%%%%%%%%%%%%%%%%
In the continuous burger example,
the true reverse-time dynamics reflect
the structure of the training weight distribution.
Recall that, 
for our given training data, 
under $p_{\mathrm{data}}(\vec{w})$,
the bun weight is 55\,g,
the patty weight 45\,g, and
the cheese weight either 14\,g or 0\,g,
with normalized coordinates
$\vec{w}_1 = [\,0,0,0\,]$ and
$\vec{w}_2 = [\,0,0,-1\,]$.
Suppose we observe a noisy burger,
$\vec{w}_t = [\, 0,\,-1,\,0 \,]$,
a cheese sandwich 
with a negative patty coordinate, 
$w^*_{\mathrm{patty}}=-1$, 
that translates into a zero patty weight, 
$w_{\mathrm{patty}}=0$\,g.
Under the true reverse-time distribution,
it is highly unlikely that a realistic previous burger
has a zero patty weight.
Therefore, the true score,
$\nabla_{\vecs{w}} \log (p_t(\vec{w}))$,
points in a direction that increases
the patty weight and moves the sample
toward the high-density region, near
$\vec{w}_t = [\,0,0,0\,]$ or
$\vec{w}_t = [\,0,0,-1\,]$.
If the learned score, $\vec{s}_{\vecs{\theta}}
(\vec{w}_t,t)$,
predicts a different direction,
the score-matching loss ${\mathcal{L}}(\theta)$ between
the true reverse-time process and
the learned reverse-time process increases.
Minimizing the score-matching loss
ensures that the reverse diffusion process
restores realistic ingredient weights,
while allowing natural continuous variability
around the learned burger manifold.} \\[6.pt]
%%%%%%%%%%%%%%%%%%%%%%%%%%%%%%%%%%%%%%%%%%%%%%%%%%%%%%%%%%%%%%%%%%%%%%%%%%
{\it{\textsf{\textbf{Geometric interpretation.}}}
%%%%%%%%%%%%%%%%%%%%%%%%%%%%%%%%%%%%%%%%%%%%%%%%%%%%%%%%%%%%%%%%%%%%%%%%%%
Forward diffusion corresponds to a probability flow 
that contracts samples toward the origin and adds isotropic Gaussian noise, 
to converge towards a spherical Gaussian distribution. 
Reverse diffusion 
introduces a learned drift term 
that follows the gradient field of the log-density, 
and steers samples 
toward regions of high probability under the data distribution. 
Geometrically, forward diffusion spreads the probability mass,
while reverse diffusion pulls it back toward the data manifold.}\\[6.pt]
%%%%%%%%%%%%%%%%%%%%%%%%%%%%%%%%%%%%%%%%%%%%%%%%%%%%%%%%%%%%%%%%%%%%%%%%%%
{\it{\textsf{\textbf{Analogy to mechanics.}}}
%%%%%%%%%%%%%%%%%%%%%%%%%%%%%%%%%%%%%%%%%%%%%%%%%%%%%%%%%%%%%%%%%%%%%%%%%%
From a mechanics perspective, 
the reverse process defines time-reversed stochastic dynamics, 
in which the score function, $\nabla_{\vecs{w}} \log (p_t(\vec{w}))$, 
acts as an effective force 
that drives the system toward high-probability configurations. 
This is analogous to gradient flow in an energy landscape, 
where the log-density plays the role of a free energy. 
The resulting dynamics 
resemble rare-event sampling and barrier-crossing processes
such as Kramers escape problems \cite{kramers1940}, 
and connect to probabilistic path formulations 
of stochastic dynamics in the spirit of Onsager's prinicple \cite{onsager1953}.}
%%%%%%%%%%%%%%%%%%%%%%%%%%%%%%%%%%%%%%%%%%%%%%%%%%%%%%%%%%%%%%%%%%%%%%%%
\begin{figure}[h]
\centering
\includegraphics[width=1.00\textwidth]{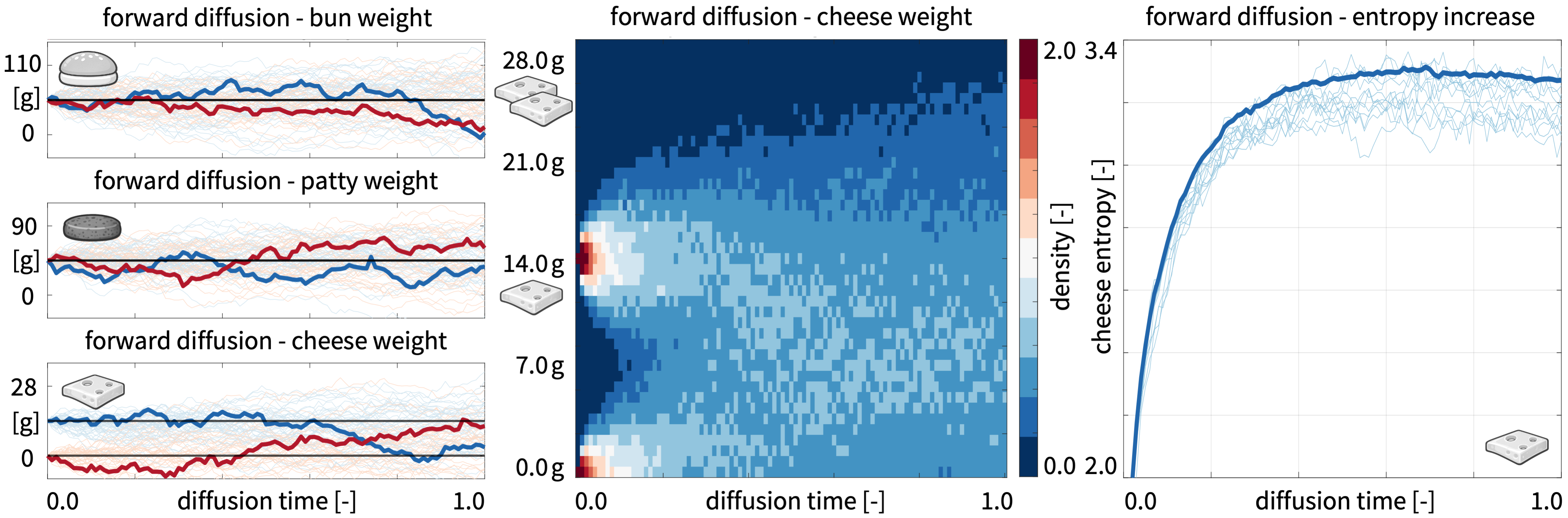} 
\caption{{\bf{\sffamily{Forward diffusion in continuous ingredient space.}}} 
Forward diffusion creates a noised state from the training data. 
Starting from two training states, cheeseburger and hamburger, 
individual trajectories 
progressively diffuse and spread in the weight space 
as reflected by the
evolution of ingredient weights (left), 
broadening of the cheese probability density (center), and 
increase in cheese entropy (right).}
\label{fig07}
%\end{figure} %\\[6.pt]
%%%%%%%%%%%%%%%%%%%%%%%%%%%%%%%%%%%%%%%%%%%%%%%%%%%%%%%%%%%%%%%%%%%%%%%%
\vspace*{0.4cm}
%%%%%%%%%%%%%%%%%%%%%%%%%%%%%%%%%%%%%%%%%%%%%%%%%%%%%%%%%%%%%%%%%%%%%%%%
%\begin{figure}[h]
\centering
\includegraphics[width=1.00\textwidth]{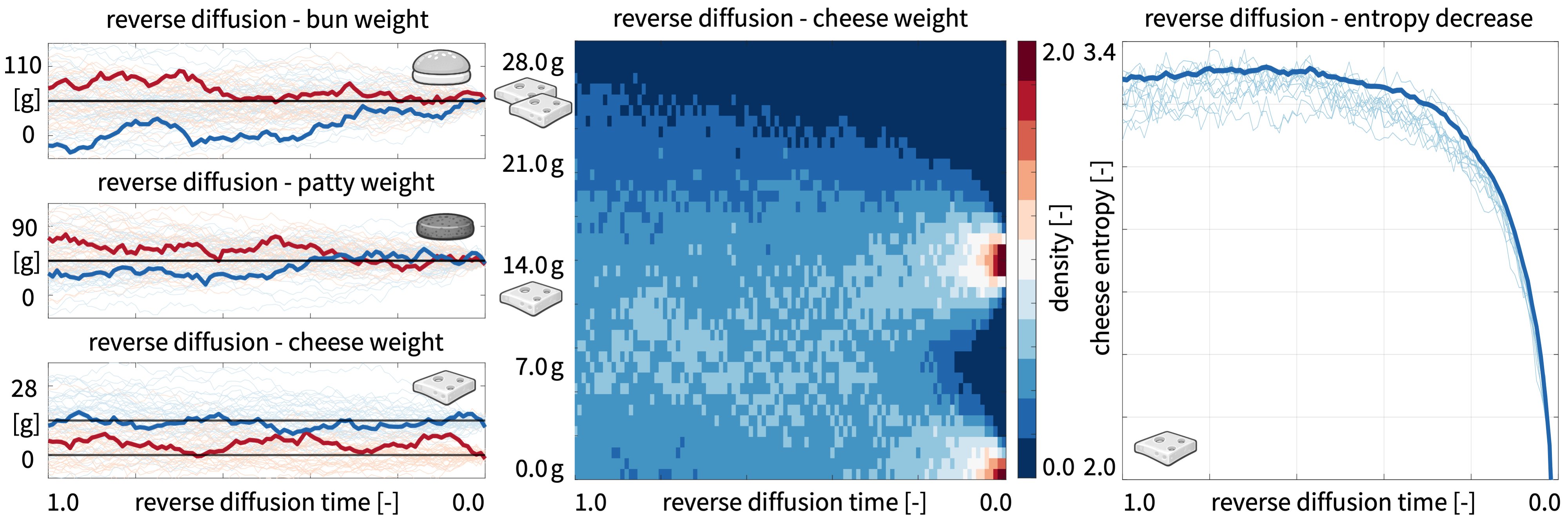} 
\caption{{\bf{\sffamily{Reverse diffusion in continuous ingredient space.}}} 
Reverse diffusion reconstructs the training distribution from the noised state. 
Starting from a diffuse distribution, 
trajectories progressively concentrate back onto the original two burgers 
as reflected by the
convergence of ingredient weights (left), 
re-localization of the probability density (center), and 
decrease in cheese entropy (right). 
Thin lines indicate stochastic realizations; 
thick lines show ensemble averages.}
\label{fig08}
\end{figure} %\\[6.pt]
%%%%%%%%%%%%%%%%%%%%%%%%%%%%%%%%%%%%%%%%%%%%%%%%%%%%%%%%%%%%%%%%%%%%%%%%
%%%%%%%%%%%%%%%%%%%%%%%%%%%%%%%%%%%%%%%%%%%%%%%%%%%%%%%%%%%%%%%%%%%%%%%%
\begin{figure}[h]
\centering
\includegraphics[width=1.00\textwidth]{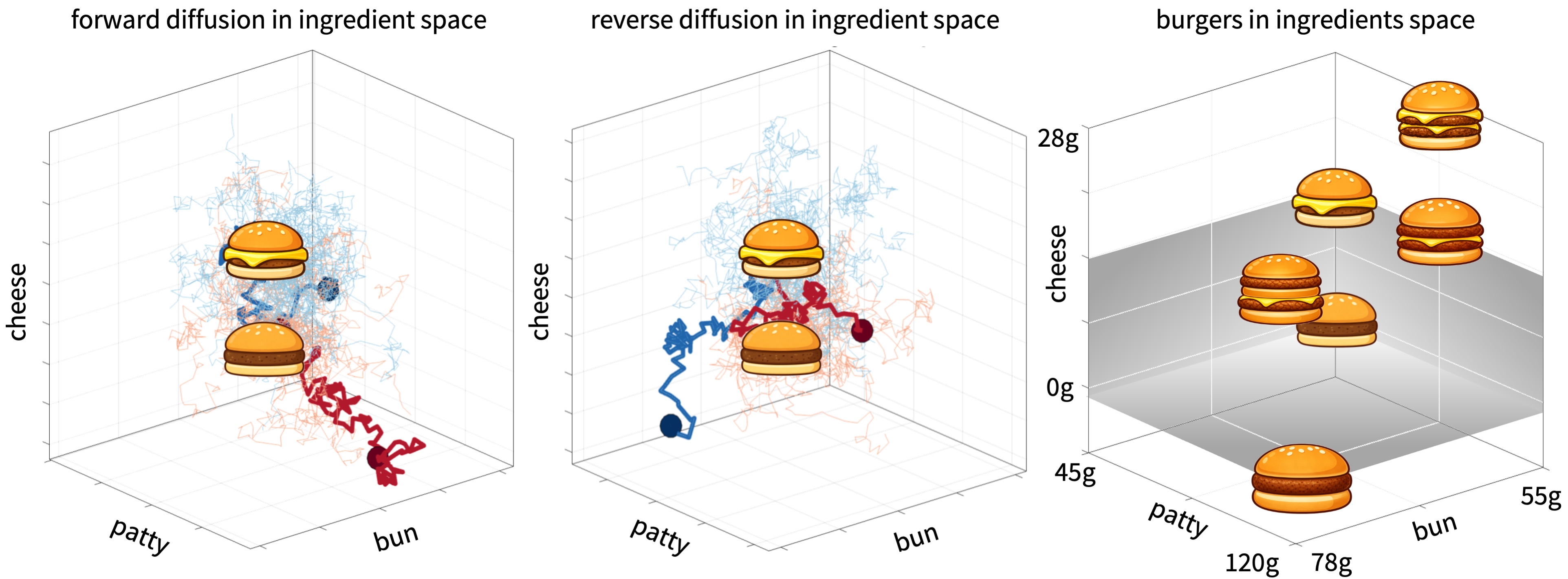} 
\caption{{\bf{\sffamily{Forward and reverse diffusion in continuous ingredient space.}}} 
Forward diffusion creates the noised state from the training data, Hamburger (red lines) and Cheeseburger (blue lines) (left).
Reverse diffusion reconstructs the training data, Hamburger (red lines) and Cheeseburger (blue lines), from the noised state (middle). 
Illustration of the training data, Hamburger and Cheeseburger,
and other possible burgers,
Mc Double, Big\,Mac, Double Cheeseburger, and Quarter Pounder,
in ingredient weight space (right).}
\label{fig09}
\end{figure} %\\[6.pt]
%%%%%%%%%%%%%%%%%%%%%%%%%%%%%%%%%%%%%%%%%%%%%%%%%%%%%%%%%%%%%%%%%%%%%%%%
%%%%%%%%%%%%%%%%%%%%%%%%%%%%%%%%%%%%%%%%%%%%%%%%%%%%%%%%%%%%%%%%%%%%%%%%%%
\subsection{\textsf{\textbf{Continuous forward and reverse diffusion}}}
%%%%%%%%%%%%%%%%%%%%%%%%%%%%%%%%%%%%%%%%%%%%%%%%%%%%%%%%%%%%%%%%%%%%%%%%%%
\noindent
We now illustrate the behavior of continuous diffusion in the three-dimensional ingredient space defined by bun, patty, and cheese weights 
(Figs. \ref{fig07} and \ref{fig08}).
In analogy to the discrete diffusion example 
(Figs. \ref{fig04} and \ref{fig05}), 
we adopt a constant diffusion rate,
$\beta(t)=\beta$,
for which the integrated noise simplifies to
$\alpha(t)=\beta t$,
the signal attenuation factor is
$\mu(t)=e^{-\beta t/2}$,
and the noise variance is
$\sigma(t) = (1-e^{- \beta t})$, 
with a time $T$=1 and 100 steps.
%%%
During {\it{forward diffusion}} (Fig. \ref{fig07}), 
the model begins with a distribution concentrated on two equally probable training burgers, cheeseburger and hamburger, whose ingredient weights define two distinct points in the continuous space. As diffusion proceeds, stochastic trajectories spread outward from these initial states, leading to increasing variability in bun, patty, and cheese weights (left). This spreading is reflected in the progressive broadening of the probability density in the cheese space (center), where initially sharp peaks evolve into a diffuse distribution. Over time, the distribution approaches a Gaussian-like equilibrium with maximal entropy, as evidenced by the convergence of the entropy curves toward a plateau (right). This behavior reflects the role of forward diffusion as a stochastic mixing process that continuously perturbs and ultimately erases information about the initial configuration (Fig. \ref{fig09}, left).
During {\it{reverse diffusion}} (Fig. \ref{fig08}), 
the model inverts the diffusion dynamics. 
Rather than starting from unconstrained Gaussian noise, we initialize the reverse process by sampling directly from the terminal forward distribution $p_T(\vec{w})$.
Reverse diffusion progressively contracts trajectories back toward the training states. 
This manifests itself in a convergence of ingredient weights toward the characteristic values of the cheeseburger and hamburger (left), accompanied by a re-localization of probability density into concentrated regions in the cheese space (center). The entropy decreases toward its initial value (right) as the burgers recover structured information (Fig. \ref{fig09}, middle). 
The thin lines of the individual stochastic trajectories visualize the variability, while the thick lines highlight the path of a single trajectory. 
Taken together, these results demonstrate that, in the continuous setting, diffusion operates over a continuous state space, where forward diffusion spreads probability mass into a high-dimensional Gaussian distribution and the learned reverse dynamics reconstruct the underlying data manifold.
%%%%%%%%%%%%%%%%%%%%%%%%%%%%%%%%%%%%%%%%%%%%%%%%%%%%%%%%%%%%%%%%%%%%%%%%%%
\subsection{\textsf{\textbf{Generating new burgers by continuous diffusion}}}
%%%%%%%%%%%%%%%%%%%%%%%%%%%%%%%%%%%%%%%%%%%%%%%%%%%%%%%%%%%%%%%%%%%%%%%%%%
\noindent
We quantify how well continuous diffusion generates burgers beyond the two training examples, hamburger and cheeseburger. We define a burger as \emph{discovered} when a sampled trajectory enters a tolerance box of 20\% around the target burger, with dimensions of 
$\Delta \vec{w}^* = [\,\pm11.0, \pm9.0, \pm2.8\,]$~g in bun, patty, and cheese. We estimate both the pathwise discovery probability $p_{\mathrm{path}}$, which counts whether a trajectory visits the target at {\it{any time}}, and the endpoint discovery probability 
$p_{\mathrm{end}}$, which counts whether the {\it{end state}} lands inside the tolerance box. We sample 5 million trajectories over a total time of $T=1$ with 100 steps at $\Delta t$ = 0.01 each. 
%%%%%%%%%%%%%%%%%%%%%%%%%%%%%%%%%%%%%%%%%%%%%%%%%%%%%%%%%%%%%%%%%%%%%%%%%%
\begin{table*}[t]
\centering
\caption{{\textsf{\textbf{Generating new burgers by continuous diffusion.}}}
Burgers, weights, coordinates, squared distance to training manifold, and 
probabilities $p_{\text{path}}$ and $p_{\text{end}}$ 
that a single trajectory passes through or ends at the target burger
at a 20\% tolerance 
for continuous forward diffusion with 5M sampled trajectories with 100 steps.}
\label{tab01}
\vspace*{0.2cm}
\resizebox{\textwidth}{!}{%
  \begin{tabular}{|l|ccc
                    |cc
                    |cc|}
% \toprule
  \hline
& \multicolumn{3}{|c}  {\textbf{burger geometry}} 
& \multicolumn{2}{|c} {{\textbf{diffusion rate}} $\beta${\textbf{ = 0.10}}} 
& \multicolumn{2}{|c|}{{\textbf{diffusion rate}} $\beta${\textbf{ = 0.25}}} \\ 
  \textbf{burger}
& $\vec{w}^*$
& $\vec{w}$ 
& $d$
& $p_{\text{path}}$
& $p_{\text{end}}$ %\\ \hline \hline
%& $N_{95}^{\text{path}}$
%& $N_{95}^{\text{end}}$
& $p_{\text{path}}$
& $p_{\text{end}}$ \\ \hline \hline
%& $N_{95}^{\text{path}}$
%& $N_{95}^{\text{end}}$ 

Hamburger
& [\,55, 45, 0\,]\,g
& [\,0, 0,-1\,]
& 0.000
& 0.5012912
& 0.0562541
%& 5
%& 52
& 0.5125357
& 0.0194421 \\
%& 5
%& 153 \\

Cheeseburger
& [\,55, 45, 14\,]\,g
& [\,0, 0, 0\,]
& 0.000
& 0.5028274
& 0.0572939
%& 5
%& 51
& 0.5201618
& 0.0212541 \\  \hline 
%& 5
%& 140 \\ \hline

Mc Double
& [\,55, 90, 14\,]\,g
& [\,0, 1, 0\,]
& 1.000
& 0.0016643
& 0.0005516
%& 1799
%& 5430
& 0.0146338
& 0.0025231 \\
%& 204
%& 1186 \\

Big\,Mac
& [\,78, 90, 14\,]\,g
& [\,0.418, 1, 0\,]
& 1.084
& 0.0007236   
& 0.0002504 
& 0.0094108    
& 0.0017217  \\
%&
%& \\

Double Cheeseburger
& [\,55, 90, 28\,]\,g
& [\,0, 1, 1\,]
& 1.414
& 0.0000136
& 0.0000054
%& 220273
%& 554764
& 0.0011225
& 0.0002582 \\
%& 2675
%& 11601 \\

Quarter Pounder
& [\,72, 120, 14\,]\,g
& [\,0.309, 1.667, 0\,]
& 1.695
& 0.0000002
& 0.0000002
%& 14978660
%& 14978660
& 0.0001581
& 0.0000438 \\ \hline
%& 18959
%& 68395 \\ \hline
%\bottomrule
\end{tabular}
}
\end{table*}
%%%%%%%%%%%%%%%%%%%%%%%%%%%%%%%%%%%%%%%%%%%%%%%%%%%%%%%%%%%%%%%%%%%%%%%%%%
We report the results for six burgers in both raw weight space $\vec{w}^*$ and transformed cheeseburger-centered coordinates $\vec{w}$ (Tab.~\ref{tab01}).
The training data define a one-dimensional manifold, the line segment connecting hamburger and cheeseburger. The burgers we seek to discover are located progressively further away from this manifold: McDouble at a distance $d$=1, Big\,Mac at $d$=1.084, Double Cheeseburger at $d$=1.414, and Quarter Pounder at $d$=1.695.  
The probabilities $p_{\text{path}}$ and $p_{\text{end}}$ 
quantify the likelihood that a single trajectory passes through or ends at a defined target burger.
%%%%%%%%%%%%%%%%%%%%%%%%%%%%%%%%%%%%%%%%%%%%%%%%%%%%%%%%%%%%%%%%%%%%%%%%
\begin{figure}[t]
\centering
\includegraphics[width=0.62\textwidth]{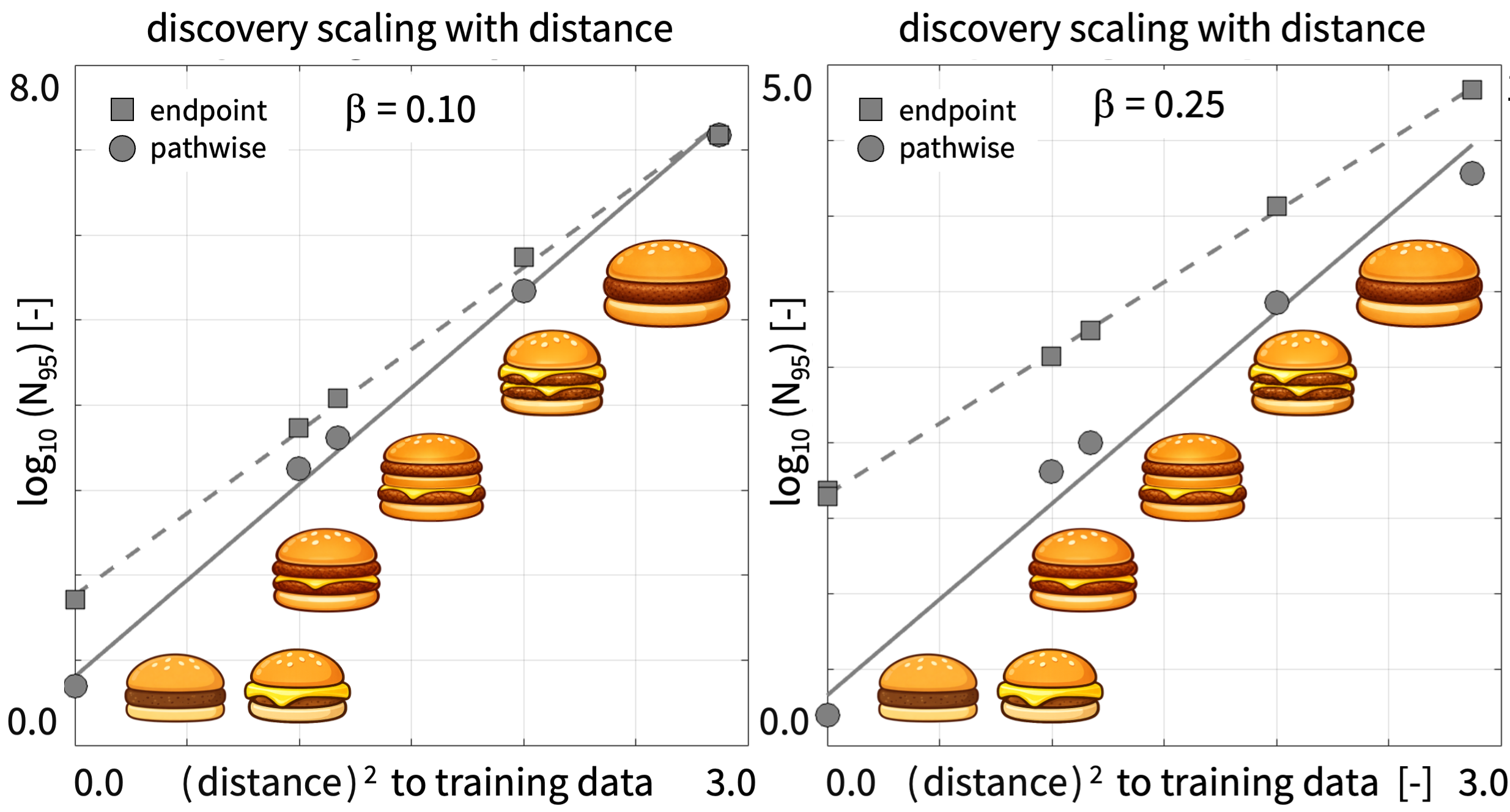} 
\caption{{\bf{\sffamily{Generating new burgers by continuous diffusion.}}}
Sampling complexity for discovering new burgers,
Mc Double, Big\,Mac, Double Cheeseburger, and Quarter Pounder,
beyond the training data, 
Hamburger and Cheeseburger. 
Number of sample trajectories required for 95\% discovery vs.
squared distance to training manifold
for diffusion rates $\beta$=0.10 (left) and $\beta$=0.25 (right).
Squares and dashed lines indicate endpoint discovery;
circles and solid lines indicate pathwise discovery 
for paths with 100 steps.
Discovery scales linearly with 
the distance squared, on the logarithmic scale, 
implying an exponential growth of discovery cost 
with squared distance from training data.}
\label{fig10}
\end{figure} %\\[6.pt]
%%%%%%%%%%%%%%%%%%%%%%%%%%%%%%%%%%%%%%%%%%%%%%%%%%%%%%%%%%%%%%%%%%%%%%%%
Discovery difficulty grows sharply with distance from the training manifold (Fig. \ref{fig10}). 
Specifically,
$N_{95}=[\log(0.05)/\log(1-p)]$ 
denotes the number of independent samples required to achieve 95\% probability of discovery.
Its logarithm, $\log_{10}(N_{95})$, is approximately linear in the distance squared, $d^2$, which implies an approximately exponential increase in the number of required samples with squared geometric distance from the training data. 
For a diffusion rate of $\beta$=0.10, 
the slopes of this linear relation, visualized through the dashed and solid lines
are 1.928 and 2.263 for endpoint and pathwise discovery (Fig. \ref{fig10}, left). 
Increasing the diffusion rate to $\beta$=0.25 notably decreases these slopes to
%$\log_{10}(N_{95}^{\mathrm{path}})=1.266\,d^2+0.831$ and
%$\log_{10}(N_{95}^{\mathrm{end}}) =0.935\,d^2+2.159$,
0.935 and 1.266 (Fig. \ref{fig10}, right),
indicating a broader exploration of the ingredient-weight space 
under stronger diffusion. \\[6.pt]
%%%%%%%%%%%%%%%%%%%%%%%%%%%%%%%%%%%%%%%%%%%%%%%%%%%%%%%%%%%%%%%%%%%%%%%%%%
{\it{\textsf{\textbf{Analogy to mechanics.}}}
%%%%%%%%%%%%%%%%%%%%%%%%%%%%%%%%%%%%%%%%%%%%%%%%%%%%%%%%%%%%%%%%%%%%%%%%%%
The observed exponential growth of sampling cost 
with squared distance is analogous to rare-event dynamics in stochastic systems, 
where transition probabilities decay exponentially with an effective energy barrier. 
In this interpretation, the squared distance from the training manifold 
plays the role of a barrier height, 
and discovering new burgers corresponds to a first-passage event across this barrier. 
This behavior is consistent with 
{\it{Arrhenius-type scaling}} \cite{arrhenius1889} 
in which transition rates depend exponentially on barrier height, 
$N_{95} \sim \exp \, (c\, d^2 \,)$,
where $c$ is a constant that depends on the diffusion rate $\beta$.} \\[6.pt]
%%%%%%%%%%%%%%%%%%%%%%%%%%%%%%%%%%%%%%%%%%%%%%%%%%%%%%%%%%%%%%%%%%%%%%%%%%
The individual burgers follow this trend: 
At small diffusion, $\beta$=0.10, the Mc Double, which differs from cheeseburger only through a second patty, requires $N_{95}^{\mathrm{path}}$=1,799 and $N_{95}^{\mathrm{end}}$=5,430 samples for pathwise and endpoint discovery. The Big\,Mac, which combines an increased bun weight with a second patty, is slightly farther from the training manifold and requires 4,139 and 11,963 samples. The Double Cheeseburger, which moves simultaneously in patty and cheese, is markedly harder to discover, with 220,273 and 554,764 samples. The Quarter Pounder is furthest from the training manifold and becomes effectively unreachable at $\beta$=0.10, as it requires approximately 1.5$\times$10$^7$ samples for pathwise and endpoint discovery.
Increasing the diffusion rate to $\beta$=0.25 substantially lowers the discovery cost of unseen burgers: The Mc Double drops from  
$N_{95}^{\mathrm{path}}$=1,799 to 204, 
the Big\,Mac from 4,139 to 317, 
the Double Cheeseburger from 220,273 to 2,675, and 
the Quarter Pounder from 1.5$\times$10$^7$ to 18,959 for pathwise discovery.
Endpoint discovery remains consistently more expensive than pathwise discovery, since it requires the {\it{final}} state rather than {\it{any}} intermediate state to reach the target. 
At the same time, stronger diffusion reduces endpoint retention of the training burgers themselves, and increases the $N_{95}^{\mathrm{end}}$ of the hamburger and cheeseburger from 50 to 150. 
This reveals a trade-off: 
Increasing the diffusion rate $\beta$ improves the exploration of unseen burgers, 
but reduces the probability of ending near the original training modes.
Taken together, this example shows that continuous diffusion does not generate new burgers uniformly. Instead, it preferentially explores a neighborhood around the training manifold, and the cost of discovering new burgers grows rapidly with distance in transformed ingredient-weight space. \\[6.pt]
%%%%%%%%%%%%%%%%%%%%%%%%%%%%%%%%%%%%%%%%%%%%%%%%%%%%%%%%%%%%%%%%%%%%%%%%%%
{\it{\textsf{\textbf{Analogy to mechanics.}}}
%%%%%%%%%%%%%%%%%%%%%%%%%%%%%%%%%%%%%%%%%%%%%%%%%%%%%%%%%%%%%%%%%%%%%%%%%%
Increasing the diffusion rate $\beta$ reduces the characteristic relaxation time and allows trajectories to explore larger regions of the state space within a fixed time horizon. 
This is analogous to increasing the temperature in Langevin dynamics \cite{langevin1908}, where stronger stochastic forcing accelerates relaxation and broadens the distribution around equilibrium. 
In both cases, higher noise levels enhance exploration while reducing retention near equilibrium configurations.} 
%%%%%%%%%%%%%%%%%%%%%%%%%%%%%%%%%%%%%%%%%%%%%%%%%%%%%%%%%%%%%%%%%%%%%%%%%%%%%%
\section{\textsf{\textbf{Generative AI for burgers}}}\label{sec04}
%%%%%%%%%%%%%%%%%%%%%%%%%%%%%%%%%%%%%%%%%%%%%%%%%%%%%%%%%%%%%%%%%%%%%%%%%%%%%%
\noindent
We now extend the discrete and continuous diffusion formulations 
from the three-ingredient benchmark with 
$n = 3$ ingredients, two training burgers, and $2^3 = 8$ possible burgers
to a real-world problem with 
$n = 146$ ingredients and 2,260 training burgers. 
Similar to Sections \ref{sec02} and \ref{sec03}, 
we represent each burger through 
a binary ingredient mask $\vec{x}$
that indicates the presence or absence of each ingredient and
the ingredient weight $\vec{w}$ that specifies the weight in grams,
\beq
\vec{x} \in \{0,1\}^{n}
\quad \mbox{and} \quad
\vec{w} \in {\mathbb{R}}^{n} \,.
\eeq
Notably, the number of possible burgers grows exponentially with the number of ingredients as $2^{n}$, 
which renders exhaustive exploration infeasible and motivates generative modeling.
For 146 independent ingredients that can be present or absent, 
the discrete design space consists of 
$
2^{146} = 8.92 \times 10^{43}
$
possible burgers.
To generate candidate burgers, 
we integrate 
a \emph{discrete diffusion model}\,\cite{ranzato2021} 
that generates ingredient masks with 
a \emph{continuous diffusion model}\,\cite{song2021} 
that generates ingredient weights,
conditional on a given mask.
Yet, the transition from the three-ingredient benchmark to the real-world problem marks a fundamental shift in the nature of the problem:
in the {\it{low-dimensional setting}} of Sections \ref{sec02} and \ref{sec03}, 
the reverse dynamics are analytically tractable, 
the score function is known, 
the reverse transitions are computable, and 
the underlying physics admit a closed-form description. 
In contrast, 
in the {\it{high-dimensional setting}}, 
the score function becomes intractable, 
the reverse dynamics can no longer be computed explicitly, and 
the physics must be inferred from data through learned models.
%%%%%%%%%%%%%%%%%%%%%%%%%%%%%%%%%%%%%%%%%%%%%%%%%%%%%%%%%%%%%%%%%%%%%%%%%%%%%%
\subsection{\textsf{\textbf{Discrete diffusion}}}
\label{sec41}
%%%%%%%%%%%%%%%%%%%%%%%%%%%%%%%%%%%%%%%%%%%%%%%%%%%%%%%%%%%%%%%%%%%%%%%%%%%%%%
\noindent
We model ingredient selection using a multinomial diffusion process 
\cite{ranzato2021},
that directly extends the discrete diffusion formulation introduced in Section \ref{sec02} to a high-dimensional setting.
We represent 
ingredient presence through a binary vector,
$\vec{x} \in \{0,1\}^{n}$,
which we modulate via forward and reverse diffusion.
The {\it{forward diffusion}} process gradually destroys structure as
\beq
p(\vec{x}_t \, | \, \vec{x}_{t-1}) 
= \mathcal{C}\!\left(
  \vec{x}_t \,\middle|\,
  [\, 1-\beta_t \,] \vec{x}_{t-1} + \beta_t/K
  \right),
\label{eq_categorical}
\eeq
where $\mathcal{C}$ denotes a {\it{categorical distribution}} over $K$ categories applied independently to each ingredient.
Its parameter is
$([\, 1-\beta_t \,] \vec{x}_{t-1} + \beta_t/K)$, 
where 
$\beta_t$ is the flip probability at time $t$, and 
$K$ is the number of categories.
%%%
In our application, ingredient selection is \emph{binary}, $K=2$, 
meaning an ingredient is either present or absent.
We can reparameterize equation (\ref{eq_categorical})
to obtain a Bernoulli distribution with parameter
$([\, 1-\beta_t \,] \, \vec{x}_{t-1} 
      + \beta_t \, [\, 1-\vec{x}_{t-1}\,])$,
which flips the ingredient state 
with a probability $\beta_t$ 
and keeps it the same with a probability $\beta_t$.
Since these distributions form a Markov chain
with independent Bernoulli corruption, 
\beq 
  p (\vec{x}_t \,|\, \vec{x}_{t-1})
= \mbox{$\prod_{i=1}^{n}$} \;
  p (x_{t,i} \,| \, x_{t-1,i})
  \quad \mbox{with} \quad
  p ({x}_{t,i} = 1 \,|\, {x}_{t-1,i})
= [\,1-\beta_t \,] x_{t-1,i} 
+ \mbox{$\frac{1}{2}$}\,\beta_t \,,
\eeq
we can explicitly calculate the distribution at any time $t$ 
in terms of the initial data $\vec{x}_0$ and the 
cumulative retention factor $\bar{\alpha}_t$,
\beq
   p(\vec{x}_t \, | \, \vec{x}_0) 
= \mathcal{C}(\vec{x}_t \, | \, 
  \bar{\alpha}_t \, \vec{x}_0 
+ \mbox{$\frac{1}{2}$} \, [\, 1-\bar{\alpha}_t \,])
  \quad \mbox{with} \quad
  \bar{\alpha}_t = \mbox{$\prod_{\tau=1}^t$} [\, 1-\beta_\tau \,] \,,
\label{eq_forward_xt_x0}  
\eeq
where $\bar{\alpha}_t$ is 
the product of the per-step retention factors $[\, 1-\beta_\tau \,]$,
which quantifies how much of the original ingredient configuration
$\vec{x}_0$ remains after $t$ diffusion steps. \\[6.pt]
The {\it{reverse diffusion}} process reconstructs structure 
by approximating the time-reversed transition probabilities. 
During training, the clean configuration $\vec{x}_0$ is known, 
which allows us to evaluate the posterior distribution via Bayes' theorem,
\beq
  p(\vec{x}_{t-1} \,|\, \vec{x}_t, \vec{x}_0)
= \frac
 {p(\vec{x}_t \,|\, \vec{x}_{t-1}) \, 
  p(\vec{x}_{t-1} \,|\, \vec{x}_0)}
 {\sum_{\vecs{y} \in \{0,1\}^n} \,
  p(\vec{x}_t \,|\, \vec{y} ) \, 
  p(\vec{y}   \,|\, \vec{x}_0)}.
\label{reverse_discrete_hd}  
\eeq
The {\it{likelihood}}
$p(\vec{x}_t \,|\, \vec{x}_{t-1})$
denotes the forward diffusion transition probability, which is given by independent Bernoulli flips with probability $\beta_t$ for each ingredient. 
The {\it{prior}}
$p(\vec{x}_{t-1} \,|\, \vec{x}_0)$
represents the marginal distribution of the forward process at time $(t-1)$, obtained from the closed-form solution of the diffusion process. 
The {\it{evidence}} acts as a normalization constant 
that ensures the posterior sums to one.
It is the sum over all possible configurations 
$\vec{y} \in \{0,1\}^n$ 
of the product 
$p(\vec{x}_t \,|\, \vec{y}) \, p(\vec{y} \,|\, \vec{x}_0)$, which corresponds to the total probability of observing $\vec{x}_t$ under all possible previous states.
As such, 
this {\it{posterior}} 
combines likelihood information from the noisy state $\vec{x}_t$ 
with prior information propagated from the original configuration $\vec{x}_0$. 
Because the forward process factorizes across ingredients, 
this update operates independently for each ingredient. 
For each ingredient $i$, we obtain a Bernoulli distribution,
\beq
p(x_{t-1,i} \,|\, x_{t,i}, x_{0,i})
\propto
p(x_{t,i} \,|\, x_{t-1,i}) \, p(x_{t-1,i} \mid x_{0,i}),
\eeq
with normalization over $x_{t-1,i} \in \{0,1\}$.
While this posterior provides a closed-form expression 
for the reverse dynamics during training, 
its evaluation requires knowledge of $\vec{x}_0$ 
and is therefore not directly usable at generation time.
We therefore introduce a neural network model 
$\mu_{\vecs{\theta}}(\vec{x}_t,t)$
where $\vec{\theta}$ denotes the network parameters
to predict the clean configuration from a noisy sample 
\cite{ho2020},
\beq
  \hat{\vec{x}}_0 
= \mu_{\vecs{\theta}}(\vec{x}_t, t)
\approx \mathbb{E}[\, \vec{x}_0 \,|\, \vec{x}_t \,]
\label{discrete_denoiser}
\eeq
We use this estimate to approximate the intractable posterior by replacing the unknown $\vec{x}_0$,
\beq
p_\theta(\vec{x}_{t-1} \,|\, \vec{x}_t)
\approx
p(\vec{x}_{t-1} \,|\, \vec{x}_t, \hat{\vec{x}}_0).
\label{discrete_reverse}
\eeq
The discrete diffusion model defines the ingredient selection $\vec{x}$, 
which conditions the continuous diffusion model to generate the corresponding ingredient weights $\vec{w}$.\\[6.pt]
%%%%%%%%%%%%%%%%%%%%%%%%%%%%%%%%%%%%%%%%%%%%%%%%%%%%%%%%%%%%%%%%%%%%%%%%%%%%%%
{\it{\textsf{\textbf{Geometric interpretation.}}}
%%%%%%%%%%%%%%%%%%%%%%%%%%%%%%%%%%%%%%%%%%%%%%%%%%%%%%%%%%%%%%%%%%%%%%%%%%%%%%
The discrete diffusion process defines a random walk 
on a $146$-dimensional hypercube, 
where vertices represent ingredient configurations 
and edges correspond to single ingredient flips. 
Forward diffusion spreads probability mass locally 
and progressively disperses it across the hypercube. 
In high dimensions, however, concentration effects dominate: 
most configurations lie near the boundary, 
typical distances increase, and 
the distribution rapidly approaches a high-entropy regime 
in which structure is difficult to distinguish. 
Reverse diffusion must therefore recover structure 
from weak signals in a space 
where relevant configurations occupy only a small fraction of the domain. 
This highlights the need 
for learned generative models 
that capture correlations between ingredients 
and guide probability mass toward realistic configurations.}
%%%%%%%%%%%%%%%%%%%%%%%%%%%%%%%%%%%%%%%%%%%%%%%%%%%%%%%%%%%%%%%%%%%%%%%%%%%%%
\subsection{\sffamily{\bfseries{Continuous diffusion}}}
\label{sec42}
%%%%%%%%%%%%%%%%%%%%%%%%%%%%%%%%%%%%%%%%%%%%%%%%%%%%%%%%%%%%%%%%%%%%%%%%%%%%%
\noindent
We model ingredient quantification using a score-based diffusion process 
in which we represent ingredient weights 
similar to Section \ref{sec03}
as a vector of real-valued variables, 
$\vec{w} \in \mathbb{R}^n$,
that we modulate via forward and reverse diffusion. 
In analogy with Section \ref{sec03},
we model {\it{forward diffusion}}
using the Ornstein--Uhlenbeck process \cite{uhlenbeck1930},
\beq
  \sca{d}\vec{w}_t
=-\mbox{$\frac{1}{2}$} \, \beta(t)\, \vec{w}_t \, \sca{d}t
+ \sqrt{\beta(t)}\, \sca{d} \vec{B}_t \,,
\label{OUforward_hd}
\eeq
and {\it{reverse diffusion}} using the time-reverse Ornstein-Uhlenbeck process \cite{song2021},
\beq
 \sca{d} \vec{w}_t
=[\,
 \mbox{$\frac{1}{2}$} \, \beta(t) \, \vec{w}_t
+\beta(t) \nabla_{\vecs{w}} \log (p_t(\vec{w}_t))
\,] \sca{d}t
+ \sqrt{\beta(t)}\, \sca{d} \tilde{\vec{B}}_t,
\label{OUreverse_hd}
\eeq
where $\sca{d} \vec{w}_t$ denotes the stochastic differential 
that represents the infinitesimal random increment of the process.
Both equations share 
{\it{deterministic drift}},
forward
$-\tfrac{1}{2}\beta(t)\vec{w}_t$
or reverse  
$+\tfrac{1}{2}\beta(t)\vec{w}_t$,
and {\it{stochastic diffusion}},
$\sqrt{\beta(t)}\,\sca{d} \vec{B}_t$ 
or 
$\sqrt{\beta(t)}\,\sca{d} \tilde{\vec{B}}_t$
in terms 
of the stochastic differential of the Brownian motion 
$\sca{d} \vec{B}_t$ 
or 
$\sca{d} \tilde{\vec{B}}_t$.
In addition, 
reverse diffusion also contains the \emph{score function},
$\nabla_{\vecs{w}} (\log p_t(\mathbf{w}))$,
a vector field that
drives diffusion uphill toward higher probability regions
\cite{pidstrigach2022,tac2024}.
The time-varying diffusion rate
$\beta(t)>0$ 
controls the magnitude of both
drift and diffusion. 
It increases as time progresses, 
$\beta(t) = \beta_{\rm{min}} + t \,[\,\beta_{\rm{max}}-\beta_{\rm{min}}]$,
from an initial weak contraction and weak noise at $\beta_{\rm{min}}$ 
towards a strong contraction and strong noise at $\beta_{\rm{max}}$. 
Similar to Section \ref{sec03},
the stochastic differential equation (\ref{OUforward_hd}) 
has the following closed-form conditional distribution,
\beq
p( \vec{w}_t |\, \vec{w}_0)
=
\mathcal{N} 
(\, \vec{w}_t |\, \mu(t)\,\vec{w}_0 , \sigma(t)\vec{I} \,) \,,
\label{OUforward_solution_hd}
\eeq
with
\beq
\mu(t) = \exp (\,- \mbox{$\frac{1}{2}$}{\alpha(t)} \,)
\quad \mbox{and} \quad
\sigma(t) = 1 - \exp(-\alpha(t)) 
\quad \mbox{and} \quad
\alpha(t)= \mbox{$\int_0^t$} \; \beta(s)\,\sca{d}s \,,
\eeq
where
$\mu(t)$ is the signal attenuation factor
%that determines how much of $w_0$ remains at time $t$,
$\sigma(t)$ is the noise variance accumulated by time $t$, and
$\alpha(t)$ is the integrated noise.\\[6.pt]
%%%%%%%%%%%%%%%%%%%%%%%%%%%%%%%%%%%%%%%%%%%%%%%%%%%%%%%%%%%%%%%%%%%%%%%%%%
{\textsf{\textbf{Score function.}}}
%%%%%%%%%%%%%%%%%%%%%%%%%%%%%%%%%%%%%%%%%%%%%%%%%%%%%%%%%%%%%%%%%%%%%%%%%%
The score function,
$\nabla_{\vecs{w}} \log (p_t(\vec{w}))$,
is the gradient of the log-density at time $t$.
It points in the direction of the steepest increase in probability density.
In contrast the low-dimensional system in Section \ref{sec03},
where we could evaluate the score function explicitly, 
the high-dimensional setting makes the marginal density $p_t(\vec{w})$ intractable:
Analytically evaluating the score function 
%$\nabla_{\vecs{w}} \log p_t(\vec{w})$ 
would require integration over all training configurations, which becomes computationally prohibitive.
We therefore approximate the vector-valued score function 
by a neural network model \cite{song2021},
\beq
\vec{s}_{\vecs{\theta}}(\vec{w},t)
\approx
\nabla_{\vecs{w}} \log p_t(\vec{w})) \,,
\label{score_approx01}
\eeq
parametrized in terms of the parameters $\vec{\theta}$, 
the network weights and biases.
We train the neural networks 
to learn the score $\vec{s}_{\vecs{\theta}}(\vec{w},t)$
%conditioned on the ingredient mask $\vec{x}$
by minimizing the {\it{score-matching}} loss $\mathcal{L}$,
the error between 
the true score function, $\nabla_{\vecs{w}} \log p_t(\vec{w}))$, and 
the approximated score of the model, $\vec{s}_{\vecs{\theta}}(\vec{w},t)$,
\beq
\mathcal{L}(\vec{\theta})
=
\mbox{$\int_0^T$} \,
\mathbb{E}_{p_t}
[ \,
\|
\nabla_{\vecs{w}} \log p_t(\vec{w}))
-
\vec{s}_{\vecs{\theta}}(\vec{w},t)
\|^2 \,] \sca{d}t \,, 
\label{eq_loss_revised}
\eeq
where
$\|\cdot\|$ is the Euclidean norm,
$\mathbb{E}_{p_t}$ denotes the expectation taken over noisy samples $\vec{w}_t$,
and the integral over $t$ averages the loss across all noise levels.
The term inside the expectation,
$\| \nabla_{\vecs{w}} \log p_t(\vec{w}))
- \vec{s}_{\vecs{\theta}}(\vec{w},t) \|^2$,
penalizes discrepancies between
the true direction in which the probability density increases most, 
$\nabla_{\vecs{w}} \log p_t(\vec{w}))$, 
and the direction predicted by the model,
$\vec{s}_{\vecs{\theta}}(\vec{w},t)$.
By minimizing the loss function over the parameter space $\vec{\theta}$,
as $\vec{\theta}^\ast = \min_{\,{\vecs{\theta}}} \, \mathcal{L}(\vec{\theta})$,
we find the model parameters $\vec{\theta}^\ast$ that best
approximate the true gradient field of the log-density.\\[6.pt]
%%%%%%%%%%%%%%%%%%%%%%%%%%%%%%%%%%%%%%%%%%%%%%%%%%%%%%%%%%%%%%%%%%%%%%%%%%%%%%
{\sffamily{\bfseries{Continuous diffusion conditioned on ingredient selection.}}}
%%%%%%%%%%%%%%%%%%%%%%%%%%%%%%%%%%%%%%%%%%%%%%%%%%%%%%%%%%%%%%%%%%%%%%%%%%%%%%
Equations (\ref{OUforward_hd}) to (\ref{eq_loss}) 
present continuous diffusion modeling in its general form. 
%%%
In practice, however,
we seek to model a 
{\it{conditional probability distribution}}, 
$p(\vec{w}|\vec{x})$, 
conditioned on the recipe mask
$\vec{x}$ produced by the discrete diffusion model. 
We therefore model the conditional distribution 
$p_t(\vec{w}\,|\,\vec{x})$, 
which we induce 
by conditioning the initial data distribution 
$p(\vec{w}_0\,|\,\vec{x})$ 
and propagating it through the forward diffusion process.
This conditioning allows the model 
to generate ingredient weights 
that are consistent with a given ingredient selection 
while preserving the statistical structure of the training data.
Accordingly, we redefine the score function as
\beq
\vec{s}_{\vecs{\theta}}(\vec{x},\vec{w},t)
\approx
\nabla_{\vecs{w}} \log p_t(\vec{w}|\vec{x})) \,.
\label{score_approx_conditional}
\eeq
Now, instead of training the score-matching loss (\ref{eq_loss})
to learn the score function 
$\vec{s}_{\vecs{\theta}}(\vec{w},t)$, 
we train the revised score-matching loss,
\beq
\mathcal{L}(\vec{\theta})
=
\mbox{$\int_0^T$} \,
\mathbb{E}_{p_t}
[ \,
\|
\nabla_{\vecs{w}} \log p_t(\vec{w}\,|\,\vec{x}))
-
\vec{s}_{\vecs{\theta}}(\vec{x},\vec{w},t)
\|^2 \,] \sca{d}t \,, 
\label{eq_loss}
\eeq
to learn the revised score function 
$\vec{s}_{\vecs{\theta}}(\vec{x},\vec{w},t)$
to approximate the 
gradient field of the conditional log-density,
$p_t(\vec{w}|\vec{x})$. 
Combined with the discrete diffusion model for ingredient selection, this conditional formulation enables the generation of complete burgers defined by both ingredient selection $\vec{x}$ and ingredient weights $\vec{w}$. \\[6.pt]
%%%%%%%%%%%%%%%%%%%%%%%%%%%%%%%%%%%%%%%%%%%%%%%%%%%%%%%%%%%%%%%%%%%%%%%%%%%%%
{\sffamily{\bfseries{Training and validation.}}}
%%%%%%%%%%%%%%%%%%%%%%%%%%%%%%%%%%%%%%%%%%%%%%%%%%%%%%%%%%%%%%%%%%%%%%%%%%%%%
We construct the {\it{mask model}} using a neural network with an embeddings layer with 1,000 embeddings for the time variable, and three fully connected layers of 512 neurons each to predict the logits of the mask model. We then use the logits with the softmax function to produce the raw probabilities. We use minibatching with $n=1,000$ for training and train with a learning rate of $5\cdot 10^{-4}$ using the Adam optimizer  for 100,000 epochs. 
%%%
We construct the {\it{value model}} using a feed-forward neural network with four hidden layers of 256 neurons each. The vectors for $\vec{x}$ and $\vec{w}$ form the input for $s_{\vecs{\theta}}$. We use a batch size of 400 and a learning rate of $1\cdot 10^{-3}$ with the Adam optimizer and train for 20,000 epochs. %\\[6.pt]
We split the training, 80\% for training and 20\% for validation
and use Nvidia H100 and L40S GPUs.
We use 1,000 steps for the noising and denoising processes 
and select scaling parameters
$\beta_{\rm{min}} = 0.001$ and $\beta_{\rm{max}}=3$ \cite{karras2022}.
%%%%%%%%%%%%%%%%%%%%%%%%%%%%%%%%%%%%%%%%%%%%%%%%%%%%%%%%%%%%%%%%%%%%%%%%%%
\subsection{{\sffamily{\bfseries{Discrete and continuous diffusion in high-dimensional ingredient space}}}}\label{sec43}
\begin{figure}[h]
\centering
\includegraphics[width=0.70\textwidth]{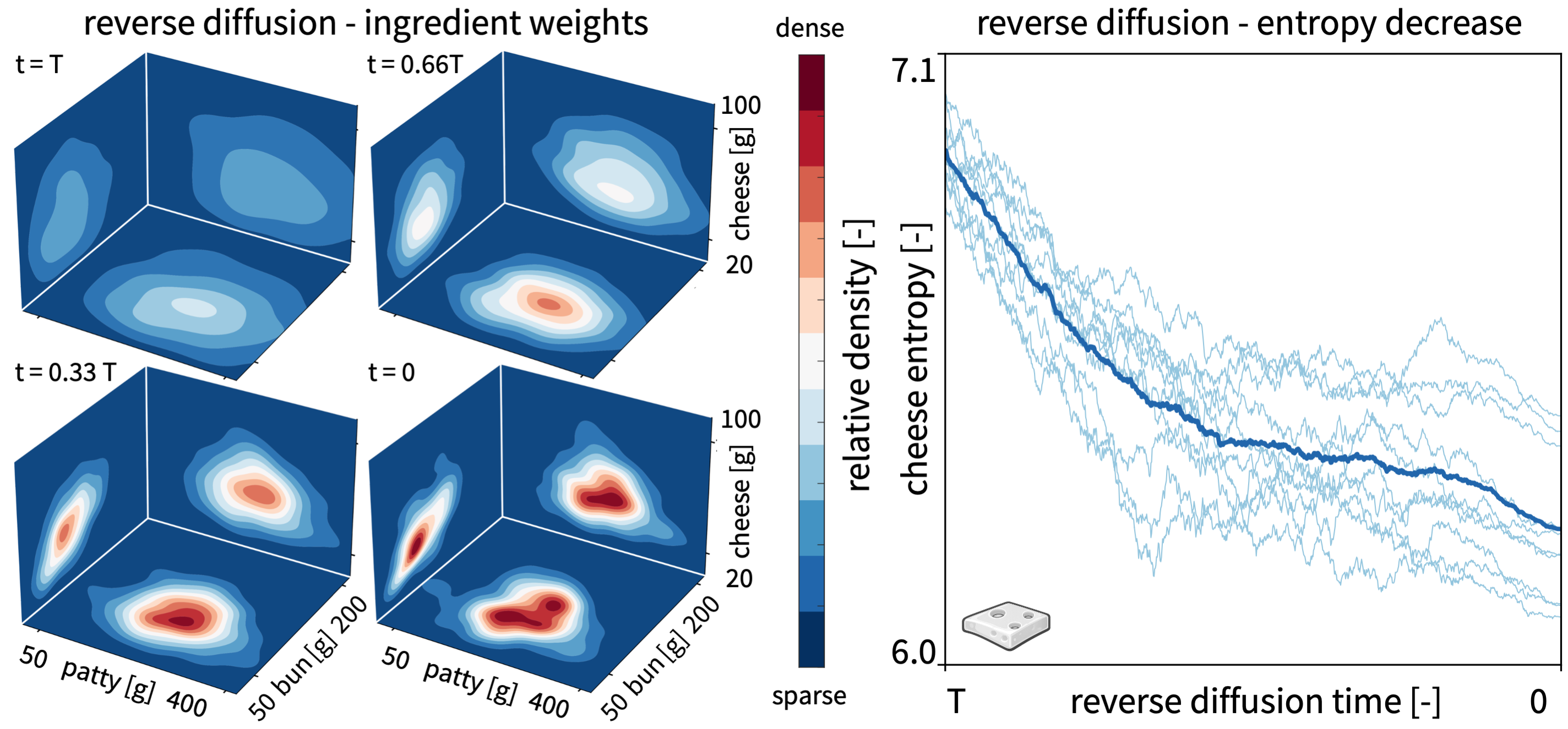} 
\caption{{\bf{\sffamily{Reverse diffusion in in high-dimensional continuous ingredient space.}}} 
Reverse diffusion recovers the training distribution from the uniform state. 
Starting from the noised state, probability mass concentrates back onto the original two burgers as reflected by 
the re-localization in probability space (left), and 
the entropy decrease (right). 
Thin lines indicate stochastic realizations; 
thick lines show ensemble averages.}
\label{fig11}
\end{figure}

%%%%%%%%%%%%%%%%%%%%%%%%%%%%%%%%%%%%%%%%%%%%%%%%%%%%%%%%%%%%%%%%%%%%%%%%%%
\noindent
We now illustrate the behavior of discrete and continuous diffusion in a high-dimensional ingredient space defined by 
$n$ = 146 ingredients (Fig. \ref{fig11}). 
In contrast to the three-ingredient benchmark with only 
2$^3$ = 8 possible burgers and two training samples, 
the present setting spans an exponentially large combinatorial design space of 
2$^{146}$ = 8.92 $\times$ 10$^{43}$ possible ingredient combinations, 
while the available training data comprise only 2,260 burgers. 
This {\it{extreme sparsity}} 
fundamentally changes the nature of the problem: 
we can no longer solve the reverse dynamics analytically;
instead, we must learn them from data.
%%%%%%%%%%%%%%%%%%%%%%%%%%%%%%%%%%%%%%%%%%%%%%%%%%%%%%%%%%%%%%%%%%%%%%%%
 %\\[6.pt]
%%%%%%%%%%%%%%%%%%%%%%%%%%%%%%%%%%%%%%%%%%%%%%%%%%%%%%%%%%%%%%%%%%%%%%%%
Figure \ref{fig11} demonstrates that, 
despite this challenge, 
diffusion retains its characteristic behavior. 
Forward diffusion progressively destroys structure 
by randomizing ingredient presence and weights, 
and drives the system toward a high-entropy state. 
Reverse diffusion--learned through neural networks--reconstructs 
structured burgers by concentrating probability mass 
onto the data manifold defined by the training set. 
The resulting stochastic trajectories 
exhibit variability at the individual level, 
while maintaining a consistent ensemble behavior
that indicates that the learned reverse dynamics 
successfully capture the underlying statistical structure 
of the high-dimensional ingredient distribution. 
Taken together, 
these results show that diffusion models 
extend naturally from low-dimensional analytical settings 
to realistic, high-dimensional design spaces, 
where they act as data-driven operators 
that transform noise into structured, physically meaningful configurations.
%%%%%%%%%%%%%%%%%%%%%%%%%%%%%%%%%%%%%%%%%%%%%%%%%%%%%%%%%%%%%%%%%%%%%%%%%%
\subsection{{\sffamily{\bfseries{Generating new burgers
in high-dimensional space}}}}
\label{sec44}
%%%%%%%%%%%%%%%%%%%%%%%%%%%%%%%%%%%%%%%%%%%%%%%%%%%%%%%%%%%%%%%%%%%%%%%%%%
\noindent
We next evaluate the ability of the learned diffusion model 
to generate new burgers beyond the 2,260 training examples 
in a space of 2$^{146}$ possible designs. 
%%%%%%%%%%%%%%%%%%%%%%%%%%%%%%%%%%%%%%%%%%%%%%%%%%%%%%%%%%%%%%%%%%%%%%%%
\begin{figure}[h]
\centering
\includegraphics[width=1.0\textwidth]{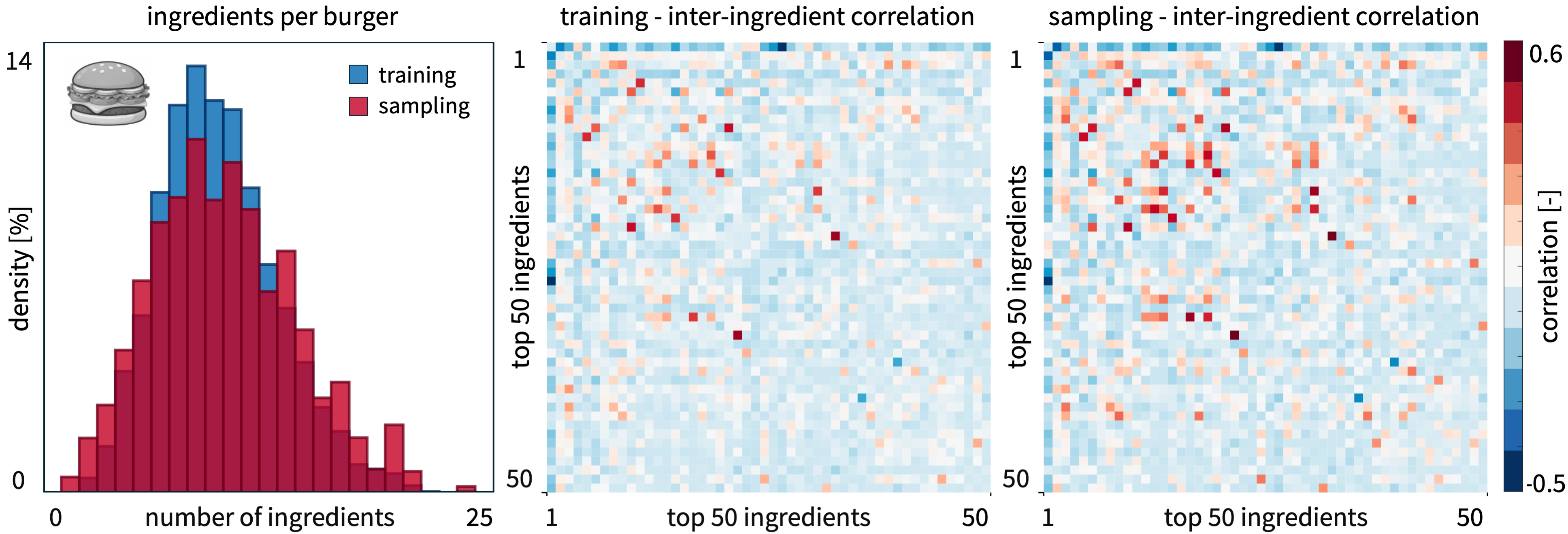} 
\caption{{\bf{\sffamily{Discrete diffusion in high-dimensional space.}}} 
Comparison of 2,260 training burgers and 1,000,000 samples generated by the learned discrete diffusion model. 
Distribution of the number of ingredients per burger (left);
inter-ingredient correlation matrix 
for the top 50 most frequent ingredients 
in the training set (middle) and in the generated samples (right).
The close agreement between training and sampling statistics 
demonstrates that the model accurately captures both 
marginal distributions and pairwise correlations in the discrete ingredient space.}
\label{fig12}
%\end{figure} %\\[6.pt]
%%%%%%%%%%%%%%%%%%%%%%%%%%%%%%%%%%%%%%%%%%%%%%%%%%%%%%%%%%%%%%%%%%%%%%%%
\vspace*{0.4cm}
%%%%%%%%%%%%%%%%%%%%%%%%%%%%%%%%%%%%%%%%%%%%%%%%%%%%%%%%%%%%%%%%%%%%%%%%
%\begin{figure}[h]
\centering
\includegraphics[width=1.0\textwidth]{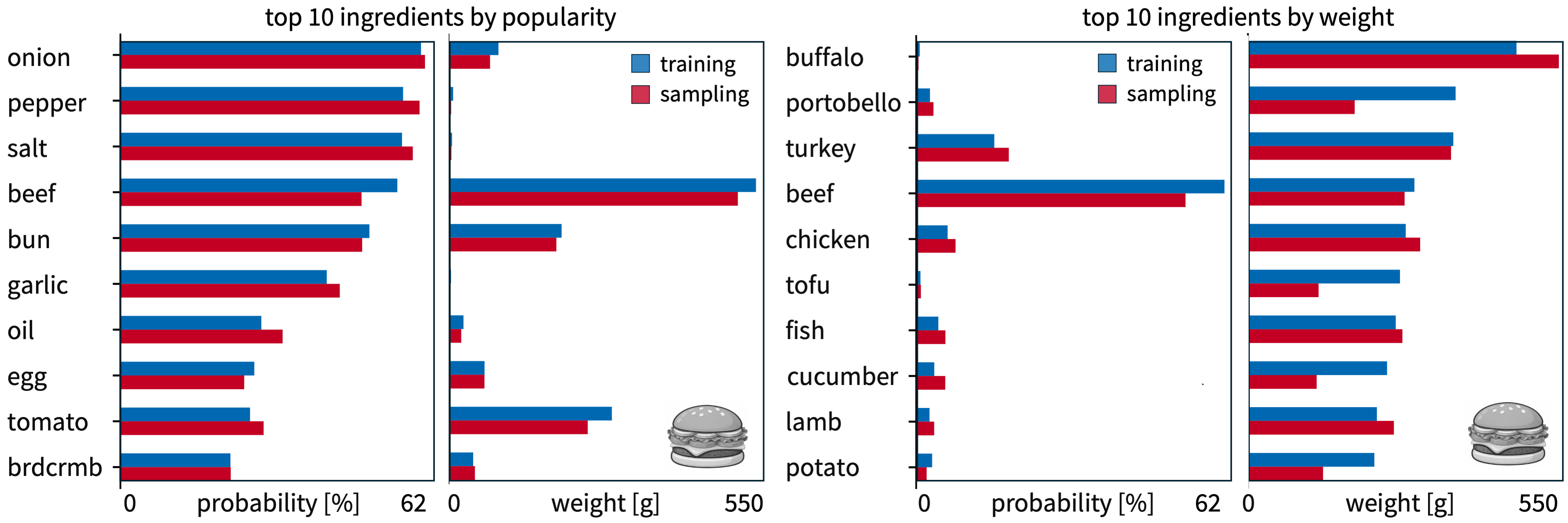} 
\caption{{\bf{\sffamily{Continuous diffusion in high-dimensional space.}}} 
Comparison of 2,260 training burgers and 1,000,000 samples generated by the learned continuous diffusion model. 
Occurrence probability of the top 10 most frequent ingredients (left) 
and average ingredient weights for the most frequent ingredients by total weight (right). 
The close agreement between training and sampling statistics 
demonstrates that the model accurately captures both 
ingredient prevalence and quantitative composition.}
\label{fig13}
\end{figure} %\\[6.pt]
%%%%%%%%%%%%%%%%%%%%%%%%%%%%%%%%%%%%%%%%%%%%%%%%%%%%%%%%%%%%%%%%%%%%%%%%
%%%%%%%%%%%%%%%%%%%%%%%%%%%%%%%%%%%%%%%%%%%%%%%%%%%%%%%%%%%%%%%%%%%%%%%%
%\vspace*{0.4cm}
%%%%%%%%%%%%%%%%%%%%%%%%%%%%%%%%%%%%%%%%%%%%%%%%%%%%%%%%%%%%%%%%%%%%%%%%
\begin{figure}[h]
\centering
\includegraphics[width=0.84\textwidth]{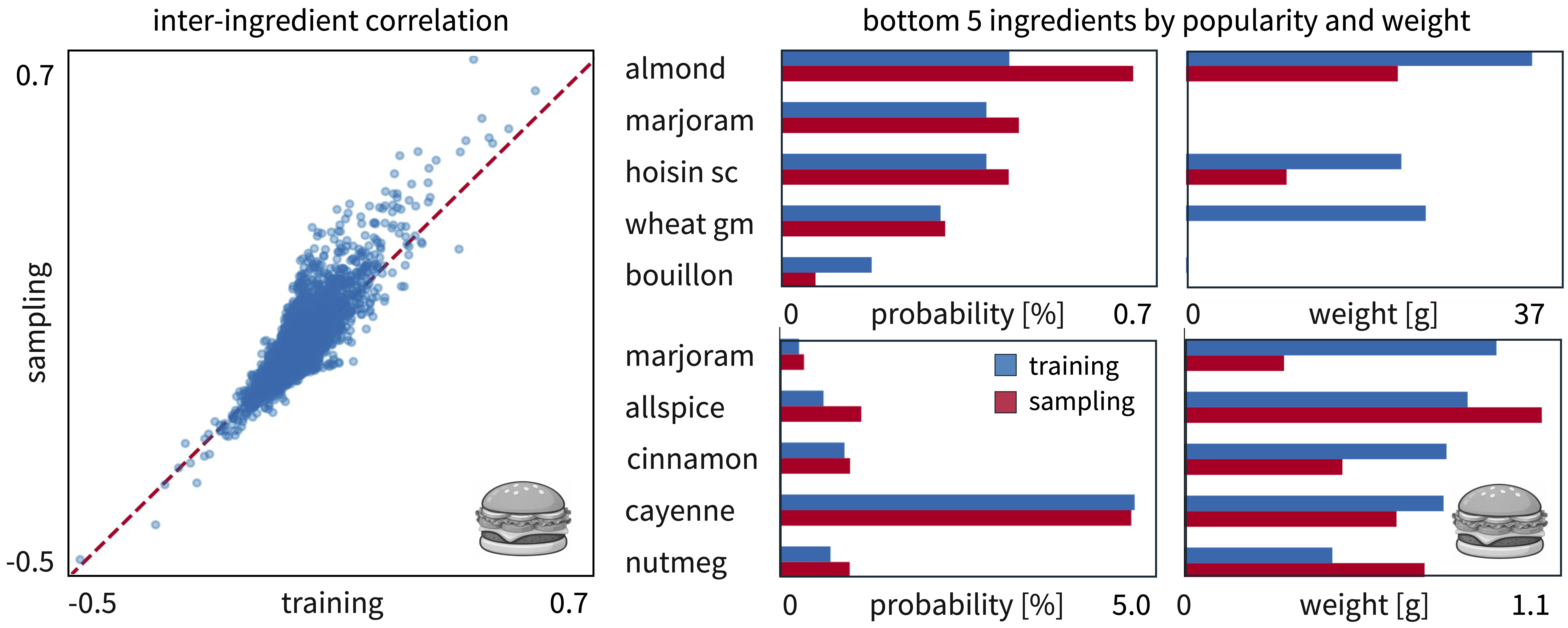} 
\caption{{\bf{\sffamily{Continuous diffusion in high-dimensional space.}}} 
Comparison of 2,260 training burgers and 1,000,000 samples generated by the learned continuous diffusion model. 
Inter-ingredient correlations between training and sampling across all 146 ingredients (left),
occurrence probability of the bottom 5 least frequent ingredients (top right) 
and average ingredient weights for the least frequent ingredients by total weight (bottom right). 
The close agreement between training and sampling statistics, 
including pairwise correlations and rare ingredients, 
demonstrates that the model accurately captures both 
high-order dependencies and rare-event structure 
across the full distribution.}
\label{fig14}
\end{figure} %\\[6.pt]
%%%%%%%%%%%%%%%%%%%%%%%%%%%%%%%%%%%%%%%%%%%%%%%%%%%%%%%%%%%%%%%%%%%%%%%%
We compare the statistical properties 
of 1,000,000 generated burgers with the training data
for both 
discrete diffusion (Fig. \ref{fig12}) 
and continuous diffusion (Figs. \ref{fig13} and \ref{fig14}).
The close agreement in 
ingredient counts, 
marginal probabilities, 
pairwise correlations, 
rare ingredients,
and weight distributions 
demonstrates that the model accurately preserves both 
low-order statistics and 
higher-order dependencies of the data. 
This includes not only the most frequent ingredients, 
but also low-frequency events and the full correlation structure across all ingredients.
This indicates that the learned generative process 
faithfully captures the underlying data distribution, 
even in an extremely high-dimensional and sparsely sampled regime.
At the same time, the generated burgers are not simple reproductions of the training set. Instead, the model samples novel combinations of ingredients and weights that lie off the observed data manifold while remaining statistically consistent with it. This is the key hallmark of generative modeling: the ability to {\it{interpolate}}, {\it{extrapolate}}, and {\it{explore}} 
unseen regions of a vast combinatorial design space.
%%%%%%%%%%%%%%%%%%%%%%%%%%%%%%%%%%%%%%%%%%%%%%%%%%%%%%%%%%%%%%%%%%%%%%%%
\begin{figure}[h]
\centering
\includegraphics[width=0.85\textwidth]{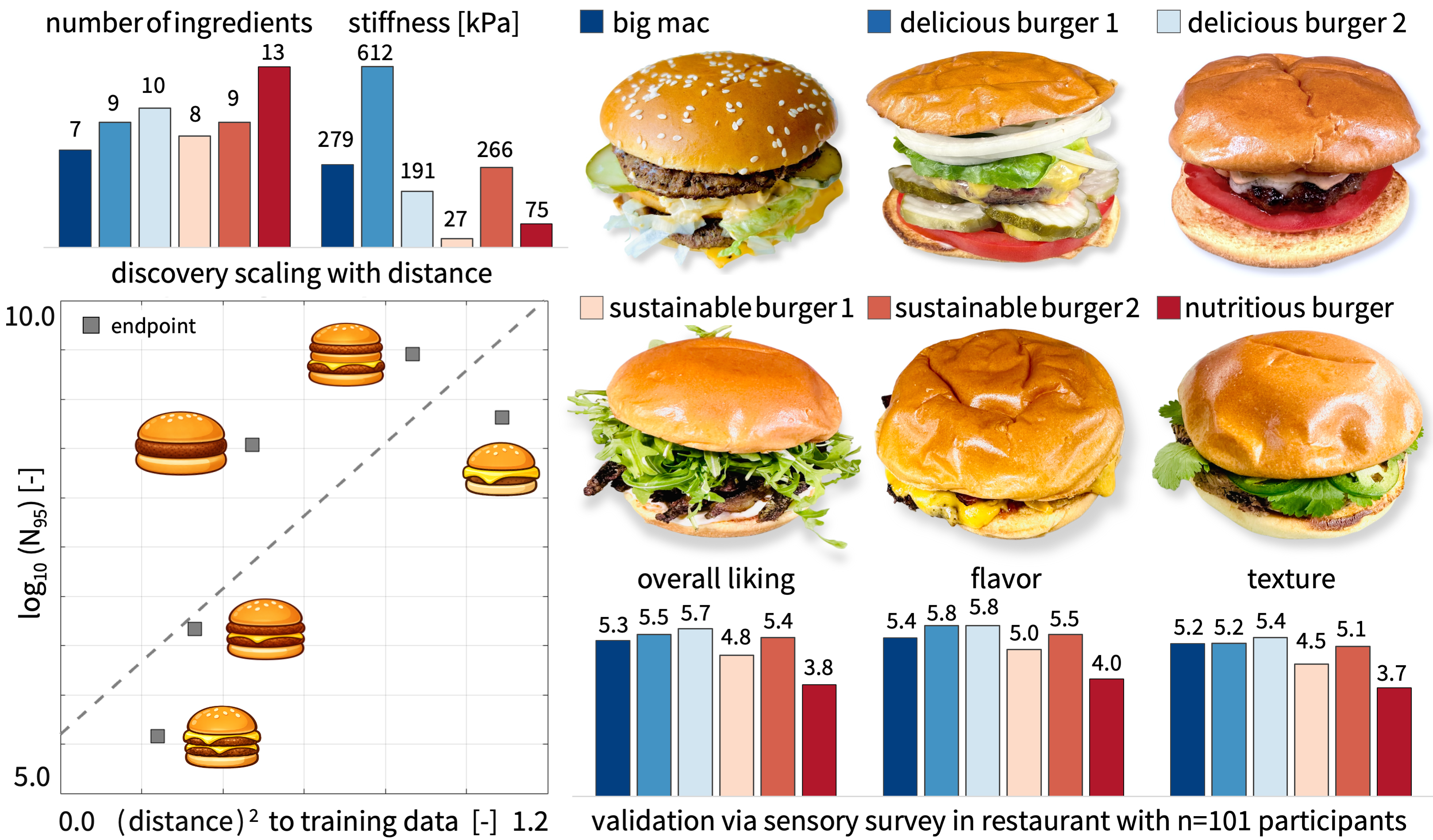} 
\caption{{\bf{\sffamily{Generating new burgers in high-dimensional space.}}} 
Sampling complexity for discovering five target burgers,
Double Cheeseburger, Mc Double, Quarter Pounder, Big\,Mac, and Cheeseburger,
reported as number of sample trajectories required 
for 95\% discovery vs. squared distance to training manifold (bottom left).
Number of ingredients and patty stiffness of
Big\,Mac and five AI-generated burgers (top left). 
Photographs of the prepared burgers, 
including the Big\,Mac, two delicious burgers, two sustainable burgers, and one nutritious burger (top right). 
Blinded sensory evaluation results 
from n = 101 restaurant participants 
rating overall liking, flavor, and texture 
on a 7-point Likert scale (bottom right). 
Three AI-generated burgers outperform the Big\,Mac in overall liking and flavor, and one AI-generated burger exceeds the Big\,Mac in texture. 
These results demonstrate that AI-driven generative design 
can produce novel food formulations 
that rival or exceed established commercial benchmarks 
in real-world sensory perception.}
\label{fig15}
\end{figure} \\[6.pt]
%%%%%%%%%%%%%%%%%%%%%%%%%%%%%%%%%%%%%%%%%%%%%%%%%%%%%%%%%%%%%%%%%%%%%%%%
From the generated samples, 
we select five AI-generated burgers
for experimental validation in a sensory study, 
using a previously reported dataset \cite{tac2026}.
While the previous study focuses on evaluating AI-generated burgers in terms of sensory performance, nutrition, and environmental impact, 
here, we use a subset of these results solely to illustrate the proposed diffusion-based framework.
We do not explicitly optimize or design these burgers in the classical sense, but sample them from the learned generative distribution, which is asymptotically dense in the high-dimensional design space and provides representative realizations of feasible configurations.
We prepare the burgers according to the AI-generated ingredient lists,
and evaluate them in a blinded sensory study with n = 101 participants
in a real restaurant setting (Fig. \ref{fig15}).  
Remarkably, 
three of the AI-generated burgers outperform the classical Big\,Mac in overall liking, 
the delicious burger 2 significantly (5.7 $\pm$ 1.2 vs.\ 5.3 $\pm$ 1.5, p$<$0.05), while 
the delicious burger 1 (5.5 $\pm$ 1.4) and 
sustainable burger 2 (5.4 $\pm$ 1.6) 
show higher mean scores without reaching statistical significance.
In flavor, 
both delicious burgers significantly outperform the Big\,Mac
(both 5.8 $\pm$ 1.3 vs.\ 5.4 $\pm$ 1.5, p$<$0.05).
Notably, the sustainable burger 2, 
which uses an animal-meat–mushroom blend, 
scores comparably to all entirely animal-based burgers across key attributes (p$>$0.05), 
including overall liking (5.4 $\pm$ 1.6), flavor (5.4 $\pm$ 1.6), and texture (5.1 $\pm$ 1.6) 
which suggests that partial substitution of animal protein with mushroom 
does not substantially degrade perceived texture \cite{stpierre2026}.
In a complementary texture profile analysis,
a double compression test that mimics the process of chewing,
the stiffnesses of the AI-generated burger patties range
from 612 $\pm$ 171\,kPa for the delicious burger 1 
to 27 $\pm$ 18\,kPa for the sustainable burger 1,
and span the range of the classical Big\,Mac
of 279 $\pm$ 130\,kPa \cite{tac2026a}.
This agreement indicates that the generated configurations 
are not only statistically consistent, 
but also physically feasible 
in terms of measurable mechanical response.
Taken together, these results demonstrate 
that diffusion-based generative models 
do not merely reproduce existing designs, 
but can discover entirely new, high-quality solutions 
that surpass established human-designed benchmarks. 
In the context of a design space of astronomical size 
and limited training data, 
this highlights the power of generative AI 
as a tool for high-dimensional discovery and data-driven innovation.
%%%%%%%%%%%%%%%%%%%%%%%%%%%%%%%%%%%%%%%%%%%%%%%%%%%%%%%%%%%%%%%%%%%%%%%%%%
\section{{\sffamily{\bfseries{Discussion}}}}\label{sec05}
%%%%%%%%%%%%%%%%%%%%%%%%%%%%%%%%%%%%%%%%%%%%%%%%%%%%%%%%%%%%%%%%%%%%%%%%%%
\noindent
The objective of this manuscript 
was to develop a unified generative framework 
that links discrete ingredient selection 
and continuous ingredient quantification 
through diffusion processes, 
and to connect diffusion-based generative AI 
to classical concepts in computational mechanics. 
We show that both formulations 
follow the same stochastic degradation 
and learned inversion principle, 
which enables the generation of novel burgers 
that preserve the structure of the training data 
while extending beyond the training manifold.
%%%%%%%%%%%%%%%%%%%%%%%%%%%%%%%%%%%%%%%%%%%%%%%%%%%%%%%%%%%%%%%%%%%

\begin{table*}[h]
\centering
\caption{\textbf{\sffamily Unified analogy of diffusion models.} Representations, directions, dimensionality, governing equations, and canonical references.}
\label{tab02}
\vspace*{0.2cm}
\renewcommand{\arraystretch}{1.05}
\begin{tabular}{|l l l|}
\hline
  \parbox{3.0cm}{\textbf{\sffamily concept}}
& \parbox{5.5cm}{\textbf{\sffamily discrete diffusion}}
& \parbox{7.5cm}{\textbf{\sffamily continuous diffusion}} \\ \hline
  example 
& ingredients
& weights\\
%%%%%%%%%%%%%%%%%%%%%%%%%%%%%%%%%%%%%%%%%%%%%%%%%%%%%%%%%%%%%%%%%%%
  state space 
& $\vec{x} \in \{0,1\}^n$ binary hypercube
& $\vec{w} \in \mathbb{R}^n$ , Euclidean space \\ \hline \hline
%%%%%%%%%%%%%%%%%%%%%%%%%%%%%%%%%%%%%%%%%%%%%%%%%%%%%%%%%%%%%%%%%%%
  {\bf{\sffamily{forward diffusion}}} 
& {\bf{\sffamily{Markov chain}}}  
& {\bf{\sffamily{Ornstein--Uhlenbeck process}}} \\ 
& Markov \cite{markov1906}, 
  Kolmogorov \cite{kolmogorov1931}
& Ornstein-Uhlenbeck \cite{uhlenbeck1930}, 
  Fokker-Planck \cite{fokker1914,planck1917} \\ \hline
%& $p(x_{t,i}=1|x_{t-1,i}) = (1-\beta_t)x_{t-1,i} + \tfrac{1}{2}\beta_t$
& $p(\vec{x}_t|\vec{x}_{t-1})$ (Eq.~\ref{eq_categorical}) 
& $\mathrm{d}\vec{w}_t 
 =-\frac{1}{2}\beta_t\vec{w}_t \sca{d}t 
  +\sqrt{\beta_t}\,\sca{d}\vec{B}_t$ (Eq.~\ref{OUforward_hd}) \\
  interpretation 
& random bit flips with flip rate $\beta_t$ 
& Gaussian perturbation with diffusion rate $\beta_t$ \\
  entropy evolution 
& probability spreads over $2^n$ states 
& density spreads toward isotropic Gaussian \\
 forward solution 
& closed form:  
\newline $p(\vec{x}_t|\vec{x}_0)$ (Eq.~\ref{evolution_hamming}) 
& \parbox{7.5cm}{closed form Gaussian:  
\newline $p(\vec{w}_t|\vec{w}_0)=\mathcal{N}(\mu(t)\vec{w}_0,\sigma(t)\ten{I})$ (Eq.~\ref{OUforward_solution_hd})} \\ \hline \hline
%%%%%%%%%%%%%%%%%%%%%%%%%%%%%%%%%%%%%%%%%%%%%%%%%%%%%%%%%%%%%%%%%%%
  {\bf{\sffamily{reverse diffusion}}} 
& {\bf{\sffamily{Bayesian inversion}}} 
& {\bf{\sffamily{reverse-time Ornstein--Uhlenbeck process}}} \\
& Bayes \cite{bayes1763} 
& Anderson \cite{anderson1982}  \\ \hline
%& $p(x_{t-1}|x_t) \propto p(x_t|x_{t-1})\,p(x_{t-1}|x_0)$
& $p(\vec{x}_{t-1}|\vec{x}_t)$ (Eq.~\ref{reverse_discrete}) 
& \parbox{7.5cm}{$  \sca{d}\vec{w}_t 
  = [\frac{1}{2}\beta_t\vec{w}_t 
  +  \beta_t\nabla \log p_t]\sca{d}t + \sqrt{\beta_t} \sca{d}\tilde{B}_t$    
  (Eq.~\ref{OUreverse_hd})} \\
  interpretation 
& probability update on graph nodes 
& drift along score field $\nabla \log p_t$ \\
  low-dim\,reverse\,solution 
& exact transitions for finite states  
& exact score of Gaussian mixture  \\
  high-dim\,reverse\,solution  
& intractable sum over $2^{146}$ states 
& intractable sum over training data \\ \hline \hline
  {\bf{\sffamily{learning}}}
& {\bf{\sffamily{denoising\,diffusion\,probabilistic\,model}}}
& {\bf{\sffamily{score-based generative model}}} \\  
& Ho et al. \cite{ho2020}
& Song et al. \cite{song2021} \\ \hline
& $p_\theta(\vec{x}_0|\vec{x}_t)$ (Eq.~\ref{discrete_denoiser})
& $s_\theta(\vec{w},t)\approx \nabla_{\vecs{w}} \log p_t(\vec{w})$ (Eq.~\ref{score_approx01}) \\
  interpretation
& learn posterior / logits  
& learn score function  \\
%  training objective 
%& cross-entropy / KL divergence  
%\newline $\mathcal{L}_{\text{CE}}$ (Eq.~\ref{loss_cross_entropy}) 
%& Score matching  
%\newline $\mathcal{L}=\int \|\nabla \log p_t - s_\theta\|^2 dt$ (Eq.~\ref{eq_loss}) \\
  sampling 
& sequential categorical updates:  
& Euler-Maruyama integration:  \cite{maruyama1954}\\
%\; \mbox{with} \, $\boldsymbol{\xi} \sim \mathcal{N}(0,\ten{I})$\\
& $\vec{x}_{t-1}\sim p_\theta(\vec{x}_{t-1}|\vec{x}_t)$ 
& \parbox{7.5cm}{%
\noindent
\(\begin{aligned}[t]
\vec{w}_{t-\Delta t}
&= \vec{w}_t
 + [
     \frac{1}{2}\beta_t\vec{w}_t
     + \beta_t\, s_\theta(\vec{w}_t,t)
   ]\Delta t \\
&\hspace{0pt}
 + \sqrt{\beta_t\Delta t}\,\vec{\xi}
\end{aligned}\)
} \\ \hline\hline

{\bf{\sffamily{unifying view}}}
& \multicolumn{2}{l|}{{\bf{\sffamily{learning to reverse entropy-increasing stochastic processes}}}} \\ \hline 

  generative role
& select ingredients $\vec{x}$ 
& assign weights $\vec{w}$ \\ 

  geometric interpretation 
& graph transport on hypercube 
& continuum transport on energy landscape \\ %\hline \hline

  mechanics analogy 
& probability flux on discrete graph
& probability density flow in continuous field \\ %\hline \hline

  unifying principle 
& learn reverse Markov dynamics 
& learn reverse stochastic dynamics \\ \hline 
\end{tabular}
\end{table*}
\noindent {\textsf{\textbf{Analogy of discrete and continuous diffusion in low dimensions.}}}
%%%%%%%%%%%%%%%%%%%%%%%%%%%%%%%%%%%%%%%%%%%%%%%%%%%%%%%%%%%%%%%%%%%%%%%%%%
Sections \ref{sec02} and \ref{sec03} 
highlight the direct analogy between {\it{discrete}} and {\it{continuous}} diffusion 
as two parallel formulations 
of the same generative principle 
in a {\it{low-dimensional}} setting. 
In the discrete case, 
burger configurations evolve on the finite state space $\{0,1\}^3$ 
via a Markov chain (Eq.~\ref{discrete_one_step}), 
admit a closed-form solution (Eq.~\ref{evolution_hamming}), 
and can be reversed exactly using Bayes' theorem (Eq.~\ref{reverse_discrete}), 
with sampling performed through categorical transitions.
In the continuous case, 
ingredient weights evolve in $\mathbb{R}^3$ 
via an Ornstein--Uhlenbeck stochastic differential equation (Eq.~\ref{OUsmall}), 
admit a Gaussian closed-form solution (Eq.~\ref{small_gauss}), 
and are reversed through the score function (Eq.~\ref{OUreverse_small}), 
with trajectories generated via Euler--Maruyama discretization. 
In both settings, 
the three-ingredient benchmark makes
the reverse dynamics analytically tractable and
provides a unified and interpretable framework that connects 
discrete probabilistic transitions with 
continuous score-based diffusion. 
This unified view expands
across discrete and continuous representations, 
forward and reverse processes, 
and low- and high-dimensional settings (Tab. \ref{tab02}).\\[6.pt]
%%%%%%%%%%%%%%%%%%%%%%%%%%%%%%%%%%%%%%%%%%%%%%%%%%%%%%%%%%%%%%%%%%%%%%%%%%
\noindent {\textsf{\textbf{Analogy of discrete and continuous diffusion 
in high dimensions.}}}
%%%%%%%%%%%%%%%%%%%%%%%%%%%%%%%%%%%%%%%%%%%%%%%%%%%%%%%%%%%%%%%%%%%%%%%%%%
Sections \ref{sec41} and \ref{sec42} extend the analogy between 
{\it{discrete}} and {\it{continuous}} diffusion 
to {\it{high-dimensional}} generative modeling, 
where exact analytical solutions are no longer tractable 
and must be approximated through learning. 
In the discrete case, 
burger configurations evolve 
on the exponentially large state space $\{0,1\}^n$ 
via a factorized Markov chain (Eq.~\ref{eq_categorical}), 
where forward transitions remain analytically defined (Eq.~\ref{eq_forward_xt_x0}), 
but reverse transitions become intractable 
due to the summation over all possible configurations (Eq.~\ref{reverse_discrete_hd}). 
We therefore approximate
the reverse process 
by a neural network that predicts the clean configuration (Eq.~\ref{discrete_denoiser}), 
via sampling through learned categorical transitions (Eq.~\ref{discrete_reverse}). 
In the continuous case, 
ingredient weights evolve in $\mathbb{R}^n$ 
via a stochastic differential equation (Eq.~\ref{OUforward_hd}), 
where the forward process admits a closed-form Gaussian solution 
(Eq.~\ref{OUforward_solution_hd}),
but the reverse-time dynamics depend on the score function through 
the reverse stochastic differential equation (Eq.~\ref{OUreverse_hd}), 
which is unknown in high dimensions and must be learned from data. 
We therefore approximate 
the reverse process 
by a neural network that predicts the underlying data structure through the score function (Eq.~\ref{score_approx01}), 
via sampling through integration of the learned stochastic dynamics. 
In both settings, 
high dimensionality transforms analytically tractable reverse dynamics 
into learning problems, 
where neural networks approximate 
either posterior distributions in discrete space (Eq.~\ref{discrete_denoiser}) 
or score functions in continuous space (Eq.~\ref{score_approx01}). 
This establishes a unified generative framework in which 
discrete and continuous diffusion differ only in representation, 
but share the same underlying principle of learning to reverse stochastic degradation. 
This transition becomes explicit when contrasting the analytically tractable low-dimensional case with the learned high-dimensional case for both discrete and continuous diffusion (Tab.~\ref{tab02}).\\[6.pt]
%%%%%%%%%%%%%%%%%%%%%%%%%%%%%%%%%%%%%%%%%%%%%%%%%%%%%%%%%%%%%%%%%%%%%%%%%%
% combinatorics... even for less ingredients...
%146 ingredients that can be present or absent
%design space consists of 2146 = 8.92 × 1043 recipes
%89,202,980,794,122,492,566,142,873,090,593,446,023,921,664
%universe is 13.8 "billion years" = 4.3 x 1017 "s"econds old
%if all 8 billion people alive today had each cooked one new burger every second since the big bang, that would be 8 x 109 x 4.35 x 1017 = 3.48 x 1027 burgers
%3,483,200,000,000,000,000,000,000,000
%we would only have explored 3.9 × 10⁻¹⁷ of the space, or 1 out of all 25 quadrillion possible burgers.
%%%%%%%%%%%%%%%%%%%%%%%%%%%%%%%%%%%%%%%%%%%%%%%%%%%%%%%%%%%%%%%%%%%%%%%%%%
\noindent {\textsf{\textbf{From burgers to matter.}}}
%%%%%%%%%%%%%%%%%%%%%%%%%%%%%%%%%%%%%%%%%%%%%%%%%%%%%%%%%%%%%%%%%%%%%%%%%%
Our results position {\it{generative AI for burgers}} 
as a direct analogue of {\it{generative AI for matter}}, 
as recently demonstrated by MatterGen,
a generative framework 
that designs inorganic materials with targeted properties \cite{zeni2025}.  
In our framework, 
ingredients play the role of chemical elements, 
and the ingredient table mirrors the periodic table 
as a discrete basis for constructing complex systems. 
Both problems share the same combinatorial explosion: 
selecting subsets and assigning weights 
defines an astronomically large design space 
that vastly exceeds available training data. 
Diffusion models address this challenge 
by learning to generate structured configurations--recipes or crystal structures--that satisfy underlying statistical and physical constraints while remaining novel \cite{bastek2025}. 
This establishes a common paradigm where generative models enable inverse design: 
burgers can be optimized for taste, nutrition, and sustainability, 
just as 
materials can be designed for stability, functionality, and resource constraints. \\[6.pt]
%%%%%%%%%%%%%%%%%%%%%%%%%%%%%%%%%%%%%%%%%%%%%%%%%%%%%%%%%%%%%%%%%%%%%%%%%%
\noindent {\textsf{\textbf{From food science to computational mechanics.}}}
%%%%%%%%%%%%%%%%%%%%%%%%%%%%%%%%%%%%%%%%%%%%%%%%%%%%%%%%%%%%%%%%%%%%%%%%%%
We can now witness the first applications of diffusion models in the field of computational mechanics: 
solving inverse problems and uncertainty quantification in physics-based systems \cite{dasgupta2026}, 
generating of constitutive laws under physics constraints \cite{tac2024}, 
forecasting and reconstructing the dynamics of incompressible flows \cite{gao2025},
reconstructing complex microstructures in materials engineering \cite{dureth2023},
generating shell structures with targeted stress-strain responses \cite{zheng2026}, and
generating energy-absorbing metamaterials and structures \cite{wang2024}. 
In contrast to these application-specific formulations, our framework provides a unified abstraction that links {\it{discrete}} and {\it{continuous}} design spaces through a common diffusion-based transport mechanism. 
Rather than targeting a single task—such as reconstruction, inference, or constitutive generation,
we interpret diffusion models as a general operator for transforming simple priors into structured matter and connect generative AI directly to the principles of computational mechanics.
While the present study does not explicitly impose physical constraints during training, we can incorporate physical feasibility by conditioning the generative process on governing laws \cite{bastek2025} or by constructing representations that satisfy the physics a priori \cite{tac2024}.
This parallel highlights a unifying paradigm: 
generative AI acts as a data-driven engine 
for high-dimensional discovery
that transforms discrete building blocks--ingredients or elements--into optimized, functional matter across domains. 
These developments build on a common theoretical foundation 
in denoising diffusion probabilistic models \cite{ho2020}, 
score-based generative modeling \cite{song2021}, and 
automated model discovery in mechanics \cite{linka2023},
and extend 
from images \cite{rombach2022} 
to proteins \cite{jumper2021}
and inorganic materials \cite{watson2023}, 
which underscores the generality 
of diffusion-based inverse design across domains.
In both discrete and continuous settings, 
training minimizes a Kullback–Leibler divergence~\cite{kullback1951}
between forward and reverse processes, 
which quantifies a mismatch in probability flux, 
and admits an interpretation as a free-energy functional 
in the sense of variational diffusion~\cite{jordan1998}.
%%%%%%%%%%%%%%%%%%%%%%%%%%%%%%%%%%%%%%%%%%%%%%%%%%%%%%%%%%%%%%%%%%%%%%%%%%%%%%
\section{\sffamily{\bfseries{Conclusion}}}
%%%%%%%%%%%%%%%%%%%%%%%%%%%%%%%%%%%%%%%%%%%%%%%%%%%%%%%%%%%%%%%%%%%%%%%%%%%%%%
\noindent
We have introduced a unified diffusion-based framework for generative design that combines discrete ingredient selection with continuous ingredient quantification.
The formulation establishes a direct correspondence between Markov chains and stochastic differential equations, and between Bayesian inversion and score-based learning.
Both models generate structured samples that recover the training data and produce novel configurations beyond the training manifold.
Discovery probability decays rapidly with geometric distance in ingredient space, which identifies distance as a governing variable for exploration in high dimensions.
These results establish diffusion models as a principled tool for design in spaces where combinatorial complexity prohibits exhaustive search.
At its core, our framework builds on concepts that are fundamental to mechanics--stochastic processes, diffusion, and inverse problems--and recasts them as engines for generative modeling.
While we use burgers as a minimal and interpretable benchmark for structured matter, our formulation extends directly to materials, biological systems, and other high-dimensional design spaces.
This work reframes diffusion not only as a model of physical processes, but as a general mechanism for the design of matter and materials.

%%%%%%%%%%%%%%%%%%%%%%%%%%%%%%%%%%%%%%%%%%%%%%%%%%%%%%%%%%%%%%%%%%%%%%%%%
\subsection*{Acknowledgements}   
%%%%%%%%%%%%%%%%%%%%%%%%%%%%%%%%%%%%%%%%%%%%%%%%%%%%%%%%%%%%%%%%%%%%%%%%%
\noindent
We thank 
Executive Chef Justin Schneider 
for his culinary expertise in creating preparation instructions, 
Caroline Cotto from NECTAR at Food System Innovations 
for stimulating discussions,
and
Alice Wistar and Alex Weissman from Palate Insights
for performing the costumer survey.
We acknowledge 
access to the Stanford Marlowe Computing Platform for high performance computing.
This research was supported by
the Schmidt Science Fellowship 
in partnership with the Rhodes Trust to Vahidullah Tac, and by
the Stanford Bio-X Snack Grant 2025,
the Stanford SDSS Accelerator Grant 2025,
the NSF CMMI Award 2320933, and
the ERC Advanced Grant 101141626 to Ellen Kuhl.
%%%%%%%%%%%%%%%%%%%%%%%%%%%%%%%%%%%%%%%%%%%%%%%%%%%%%%%%%%%%%%%%%%%%%%%%%
\subsection*{CRediT authorship contribution statement}
%%%%%%%%%%%%%%%%%%%%%%%%%%%%%%%%%%%%%%%%%%%%%%%%%%%%%%%%%%%%%%%%%%%%%%%%%
\noindent 
VT: Conceptualization, Methodology, Software, Formal analysis, Data Curation, Investigation, Validation, Writing Original Draft, Writing Review and Editing.
EK: Conceptualization, Methodology, Software, Formal analysis, Data Curation, Investigation, Validation, Writing Original Draft, Writing Review and Editing.
%%%%%%%%%%%%%%%%%%%%%%%%%%%%%%%%%%%%%%%%%%%%%%%%%%%%%%%%%%%%%%%%%%%%%%%%%
\subsection*{Data availability}
%%%%%%%%%%%%%%%%%%%%%%%%%%%%%%%%%%%%%%%%%%%%%%%%%%%%%%%%%%%%%%%%%%%%%%%%%
\noindent
Our source code, data, and examples are available at 
https:/\!/github.com/LivingMatterLab/AI4Food.
%%%%%%%%%%%%%%%%%%%%%%%%%%%%%%%%%%%%%%%%%%%%%%%%%%%%%%%%%%%%%%%%%%%%%%%%%

%%%%%%%%%%%%%%%%%%%%%%%%%%%%%%%%%%%%%%%%%%%%%%%%%%%%%%%%%%%%%%%%%%%%%%%%%%
%%%%%%%%%%%%%%%%%%%%%%%%%%%%%%%%%%%%%%%%%%%%%%%%%%%%%%%%%%%%%%%%%%%%%%%%%%
\end{document}